\documentclass[prd,twocolumn,amsmath,amssymb,floatfix]{revtex4}
\usepackage{graphicx}
\usepackage{color}
\usepackage{dcolumn}
\usepackage{bm}

\def\p{\partial}

\def\Lie{{\cal L}}

\usepackage{color}
\usepackage{float}
\usepackage{ulem}
\definecolor{cyan}{rgb}{0,0.9,0.9}
\definecolor{orange}{rgb}{0.9,0.5,0}
\definecolor{magenta}{rgb}{1,0,1}
\definecolor{purple}{rgb}{0.8,0.4,0.8}

\begin{document}


\title{Numerical stability of the Z4c formulation of general
relativity}

\author{Zhoujian Cao${}^{1}$, David \surname{Hilditch}${}^{2}$}

\affiliation{${}^1$Institute of Applied Mathematics, Academy of
Mathematics and Systems Science, Chinese Academy of Sciences,
Beijing 100190, China, ${}^2$Theoretical Physics Institute,
University of Jena, 07743 Jena, Germany}

\date{\today}

\begin{abstract}
We study numerical stability of different approaches to the
discretization of a conformal decomposition of the Z4 formulation of
general relativity. We demonstrate that in the linear, constant
coefficient regime a novel discretization for tensors is formally
numerically stable with a method of lines time-integrator. We then
perform a full set of ``{\it apples with apples}'' tests on the
non-linear system, and thus present numerical evidence that both
the new and standard discretizations are, in some sense, numerically
stable in the non-linear regime. The results of the Z4c numerical
tests are compared with those of BSSNOK evolutions. We typically 
do not employ the Z4c constraint damping scheme and find that in 
the robust stability and gauge wave tests the Z4c evolutions result 
in lower constraint violation at the same resolution as the BSSNOK 
evolutions. In the gauge wave tests we find that the Z4c evolutions 
maintain the desired convergence factor over many more light-crossing 
times than the BSSNOK tests. The difference in the remaining tests 
is marginal.
\end{abstract}


\maketitle

\tableofcontents


\section{Introduction}


There are currently two main formulations used in the numerical
evolution of astrophysically interesting spacetimes by the methods
of numerical relativity. The first was pioneered by 
Pretorius~\cite{Pre04,Pre05} and later adopted by other numerical
relativity groups, for example~\cite{LinSchKid05,PalOlaLeh06,SziPolRez06}. 
In this approach the generalized harmonic gauge formulation of 
general relativity, GHG~\cite{Fri85,Gar01}, is employed with
black hole excision. That is, inside the black hole horizon the
numerical mesh contains a ``cut-out'' region. The generalized
harmonic formulation has a number of desirable properties. The first
is that it has a trivially wave-like principal part, which allows
the construction of boundary conditions that lead to a well-posed
initial boundary value problem that can be implemented
numerically~\cite{SziWin02,LinSchKid05,KreWin06,BabSziWin06,
MotBabSzi06,Rin06a,RuiRinSar07,RinBucSch08}. That the wave-like
nature of the formulation is inherited by the constraint subsystem
means that these boundary conditions can conveniently be made
constraint preserving, and that when numerical error causes violations 
of the constraints, this violation may propagate off of the numerical 
grid, hopefully to be harmlessly absorbed by the aforementioned
boundary conditions. The evolution system also admits a constraint
damping scheme~\cite{GunGarCal05}, which has proven
important in numerical applications to avoid constraint violating
blow-ups. The second main formulation is the combination of the
BSSNOK system, the moving puncture family of gauge conditions,
and puncture black-hole initial data~\cite{BakCenCho05,CamLouMar05}.
Using this method coordinate singularities, or punctures, are
explicitly advected across the numerical mesh. The puncture method
has proven extremely robust in the evolution of even
extreme initial data~\cite{SpeCarPre08,SpeCarPre09}. But on the
other hand, whilst well-posedness for the initial value problem of
BSSNOK with suitable members of the puncture gauge family has
been established~\cite{GunGar06} the complicated structure of the
principal part has prevented the same development of boundary
conditions for the system, although interesting progress has been
made~\cite{NunSar09}.

This status begs the question: is it possible to write down a
formulation with the strengths of both GHG and BSSNOK for numerical
applications? In~\cite{BerHil09} a natural candidate, namely a
conformal decomposition of the Z4
formulation~\cite{BonLedPal03,BonPal04,BonLedPal03a,BonLedLuq04a,AliBonBon08,
BonBonPal10,BonBon10}, Z4c was identified, and indeed found to give
favourable results over those of BSSNOK in the context of spherical
symmetry. Like that of GHG, the constraint subsystem of the Z4
formulation has a trivial wave-like principal part. This structure
is inherited by the constraint subsystem of the conformally
decomposed system provided that the additional constraints
introduced by the decomposition are explicitly imposed. This
property allows the convenient construction of constraint preserving
boundary conditions for the Z4c system~\cite{RuiHilBer10}. The Z4c
system furthermore inherits the constraint damping scheme of Z4,
which was studied in detail in~\cite{WeyBerHil11}. So far the
numerical studies of Z4c have been restricted to spherical symmetry.
Recently a variant of the conformal decomposition, CCZ4, was
performed, and for the first time numerical evolutions in three
dimensions were presented~\cite{AliBonBon11}. The difference 
between Z4c and CCZ4 is that in the conformal decomposition of 
Z4c non-principal constraint additions are discarded in such a 
way to make the resulting evolution system as close as possible 
to BSSNOK. On the other hand in CCZ4 no constraint terms are 
discarded. Therefore the two formulations share the same principal 
part, and thus the same basic PDE properties. Now that there 
is robust evidence in favor of the Z4c system in spherical
symmetry we also turn our attention to three spatial dimensions.
Since there is a wealth of experience concerning the robustness of
the GHG and BSSNOK formulations in applications, significant
evidence must be presented that an alternative formulation is 
competitive, and we therefore take a conservative approach, and 
focus here on the question of numerical stability.

In section~\ref{Section:Z4c} we present the equations of motion of
the Z4c system and the gauge conditions that we employ in both our
analytical and numerical studies. In section~\ref{Section:Analysis}
we introduce a novel discretization of tensors and show that in the
linear constant coefficient approximation Z4c, coupled to the
puncture gauge, is numerically stable with this discretization and a
method of lines time-integrator. In section~\ref{Section:AwA} we
present a complete set of ``{\it apples with apples}'' tests for the
Z4c formulation. In appendices~\ref{App:weak} and~\ref{App:Not_Sym}
we discuss hyperbolicity of the conformal decomposition without
imposition of the algebraic constraints, and symmetric hyperbolicity
of the formulation with the puncture gauge. We conclude in
section~\ref{Section:Conclusions}.


\section{The Z4c formulation}
\label{Section:Z4c}


The Z4 formulation~\cite{BonLedPal03}, with constraint
damping~\cite{GunGarCal05} replaces the Einstein field
equations~$G_{ab}=8\pi T_{ab}$ by
\begin{align}
&G_{ab}=8\pi
T_{ab}-\nabla_aZ_b-\nabla_bZ_a+g_{ab}\nabla_cZ^c\nonumber\\
&\qquad +\kappa_1[n_aZ_b+n_bZ_a+\kappa_2g_{ab}n_cZ^c],
\end{align}
where $G_{ab}=R_{ab}-\frac{1}{2}Rg_{ab}$ is the Einstein tensor
while $R_{ab}$ is the Ricci tensor, $\nabla_a$ is the covariant
derivative operator compatible with $g_{ab}$ and $Z_a$ is an
additional vector field of constraints. Under the standard $3+1$
decomposition, against a timelike unit normal vector $n^a$,
defining~$\Theta\equiv-n^aZ_a$ and $Z_i\equiv\perp^a_{\,i}Z_a$ one
finds the expressions given in~\cite{GunGarCal05}. If the spacetime 
is without boundary and does not admit a Killing
vector, it can be shown that if the constraints are satisfied in
one spacelike slice then they will all vanish all
times~\cite{BonLedPal03}. We may therefore take a free-evolution
approach to the problem. For PDEs and numerical analysis we work for
in the expanded phase space, in which the constraints may be violated.
In numerical applications the constraints are to be solved for initial
data, and their compatibility with the evolution equations means that
any violation should converge away with resolution.

From the PDEs point of view, only the principal derivative additions
can affect well-posedness of the initial value problem of the
formulation as a PDE system. In order to make a conformal
decomposition we therefore discard non-principal terms in such a way
that the resulting equations of motion will have a form
similar to those of the BSSNOK formulation when written in terms of
conformal variables. Of course we keep the constraint damping terms.
The Z4c formulation equations of motion~\cite{BerHil09,RuiHilBer10}
are
\begin{align}
\p_t \gamma_{ij}&=-2\alpha K_{ij}+\Lie_\beta\gamma_{ij},\label{eqn:g}\\
\p_tK_{ij} &=-D_iD_j\alpha+\alpha [R_{ij}+ K K_{ij} - 2 K_{ik}K^k{}_j
+2\hat{D}_{(i}Z_{j)}\nonumber\\
&-\kappa_1(1+\kappa_2)\Theta\gamma_{ij}]
+4\pi\alpha [\gamma_{ij}(S-\rho_{\textrm{\tiny{ADM}}})-2S_{ij}]\nonumber\\
&+\Lie_\beta K_{ij},\label{eqn:k}
\end{align}
for the metric and extrinsic curvature, and
\begin{align}
\p_t\Theta &= \frac{1}{2}\alpha [H + 2\hat{D}^iZ_i
-2\kappa_1(2+\kappa_2)\Theta]+\Lie_\beta\Theta,\\
\p_t Z_i&=\alpha [M_i+D_i\Theta-\kappa_1Z_i]
+\gamma^{\frac{1}{3}}Z^j\p_t[\gamma^{-\frac{1}{3}}\gamma_{ij}]\nonumber\\
&+\beta^j\hat{D}_jZ_i,
\end{align}
for the constraints $\Theta$ and~$Z_i$, where here we define
\begin{align}
\hat{D}_iZ_{j}&=\gamma^{\frac{1}{3}}\gamma_{kj}\p_i[\gamma^{-\frac{1}{3}}Z^k],
\end{align}
and use the shorthands
\begin{align}
H&=R-K_{ij}K^{ij}+K^2-16\pi\rho_{\textrm{\tiny{ADM}}},\\
M_i&=D^j[K_{ij}-\gamma_{ij}K]-8\pi S_i,
\end{align}
for the Hamiltonian and momentum constraints. Similar to the conformal
transformation made in the BSSNOK formalism, we introduce conformal
variables
\begin{align}
&\tilde{\gamma}_{ij}=\gamma^{-1/3}\gamma_{ij},
&\chi=\gamma^{-1/3},\\
&\tilde{A}_{ij}=\gamma^{-1/3}(K_{ij}-\frac{1}{3}\gamma_{ij}K),
&\hat{K}=K-2\Theta,\\
&\tilde{\Gamma}^i=2\tilde{\gamma}^{ij}Z_j+
\tilde{\gamma}^{ij}\tilde{\gamma}^{kl}\tilde{\gamma}_{jk,l},
&\tilde{\Gamma}_{\textrm{\tiny{d}}}{}^i=\tilde{\Gamma}^i{}_{jk}
\tilde{\gamma}^{jk}.
\end{align}
Under this change of variables the evolution equations become
\begin{align}
\p_t\chi&=\frac{2}{3}\chi[\alpha(\hat{K}+2\Theta)-D_i\beta^{i}],\\
\p_t\tilde\gamma_{ij}&=
- 2 \alpha \tilde A_{ij}+2\tilde\gamma_{k(i}\p_{j)}\beta^k
- \frac{2}{3} \tilde\gamma_{ij}\p_{k}\beta^k
+ \beta^k\p_k\tilde{\gamma}_{ij},\\
\p_t\hat{K}&=-D_iD^i\alpha + \alpha[\tilde{A}_{ij}\tilde{A}^{ij}
+\frac{1}{3}(\hat{K}+2\Theta)^2\nonumber\\
&+\kappa_1(1-\kappa_2)\Theta]
+4\pi\alpha[S+\rho_{\textrm{\tiny{ADM}}}]+\beta^i\p_i\hat{K},\\
\p_t\tilde{A}_{ij}& = \chi [-D_iD_j\alpha+\alpha(R_{ij}
-8\pi S_{ij})]^{\textrm{tf}}
+\alpha [(\hat{K}+2\Theta)\tilde A_{ij} \nonumber\\
&- 2\tilde A_{ik}\tilde A^k{}_j]+2\tilde A_{k(i}\p_{j)}\beta^k
-\frac{2}{3}\tilde A_{ij}\p_k\beta^k
+ \beta^k\p_k\tilde A_{ij},\\
\p_t\Theta&=\frac{1}{2}\alpha\big[R-\tilde{A}_{ij}\tilde{A}^{ij}
+\frac{2}{3}(\hat{K}+2\Theta)^2-16\pi\rho_{\textrm{\tiny{ADM}}}\nonumber\\
&-2\kappa_1(2+\kappa_2)\Theta\big]+\beta^i\p_i\Theta,\\
\p_t\tilde{\Gamma}^i&= \tilde{\gamma}^{jk}\p_j\p_k\beta^i
+\frac{1}{3}\tilde{\gamma}^{ij}\p_j\p_k\beta^k
-2\tilde{A}^{ij}\p_j\alpha\nonumber\\
&+2\alpha \big[\tilde{\Gamma}^i{}_{jk}\tilde A^{jk}
-\frac{3}{2}\tilde{A}^{ij}\p_j\ln\chi
-\frac{1}{3}\tilde{\gamma}^{ij}\p_j(2\hat{K}+\Theta)\nonumber\\
&-\kappa_1(\tilde{\Gamma}^i-\tilde{\Gamma}_{\textrm{\tiny{d}}}{}^i)
-8\pi\tilde{\gamma}^{ij}S_j
\big]+\frac{2}{3}\tilde{\Gamma}_{\textrm{\tiny{d}}}{}^i\p_j\beta^j
-\tilde{\Gamma}_{\textrm{\tiny{d}}}{}^j\p_j\beta^i\nonumber\\
&+\beta^j\p_j\tilde{\Gamma}^i,
\end{align}
where we absorb the constraint addition into the Ricci tensor
according to
\begin{align}
R_{ij} &= R^{\chi}{}_{ij} + \tilde{R}_{ij},\\
\tilde{R}^{\chi}{}_{ij} &=\frac{1}{2\chi}\tilde{D}_i\tilde{D}_j\chi
+\frac{1}{2\chi}\tilde{\gamma}_{ij}\tilde{D}^l\tilde{D}_l\chi
-\frac{1}{4\chi^2}\tilde{D}_i\chi\tilde{D}_j\chi\nonumber\\
&-\frac{3}{4\chi^2}
\tilde{\gamma}_{ij}\tilde{D}^l\chi\tilde{D}_l\chi,\\
\tilde{R}_{ij} &= -\frac{1}{2}\tilde{\gamma}^{lm}
\tilde{\gamma}_{ij,lm} +\tilde{\gamma}_{k(i}\p_{j)}
\tilde{\Gamma}^k+\tilde{\Gamma}_{\textrm{\tiny{d}}}{}^k
\tilde{\Gamma}_{(ij)k}\nonumber\\
&+\tilde{\gamma}^{lm}\left(2\tilde{\Gamma}^k{}_
{l(i}\tilde{\Gamma}_{j)km}+\tilde{\Gamma}^k{}_{im}
\tilde{\Gamma}_{klj}\right).
\end{align}
In this study we consider only vacuum spacetimes,
so $S_{ij},S_i$ and $\rho_{\textrm{\tiny{ADM}}}$ all
vanish. Note that the conformal decomposition introduces two
algebraic constraints
\begin{align}
D\equiv\ln \det\tilde{\gamma}=0,\qquad
T\equiv\gamma^{ij}\tilde{A}_{ij}=0,
\end{align}
to the system. In our numerical experiments they are imposed
explicitly after every time-step. If the constraint projection is
not performed then one is performing free-evolution in the larger
phase space in which the algebraic constraints are violated, meaning
that the PDE properties of the system must be re-evaluated. We close
the system with the puncture gauge condition, which consists of the 
Bona-M\'asso lapse~\cite{BonMas94} and the gamma-driver shift 
conditions~\cite{AlcBruDie02},
\begin{align}
\label{eqn:BM_lapse}
\p_t\alpha&=-\mu_L\alpha^2\hat{K}+\beta^i\p_i\alpha,\\
\label{eqn:Gamma_driver}
\p_t\beta^i&= \mu_S\alpha^2\tilde{\Gamma}^i-\eta \beta^i
+ \beta^j\p_j\beta^i.
\end{align}
When coupled to the puncture gauge the Z4c formulation forms a PDE 
system that is generically strongly hyperbolic. In our numerical 
applications we almost always take the ``1+log'' variant of the 
lapse condition~$\mu_L=2/\alpha$ and $\mu_S=1/\alpha^2$, although 
some tests are performed with harmonic lapse~$\mu_L=1$. The shift 
damping term $\eta$ is chosen according to the needs of the numerical
test. We also sometimes employ the harmonic shift condition
\begin{align}
\label{eqn:Harm_Shift}
\partial_t\beta^i&=\alpha^2\big[\chi\tilde{\Gamma}^i
+\frac{1}{2}\chi\tilde{\gamma}^{ij}\chi_{,j}
-\chi\tilde{\gamma}^{ij}\p_j\ln\alpha\big]
+\beta^j\partial_j\beta^i.
\end{align}


\section{Numerical Stability}
\label{Section:Analysis}


In this section we demonstrate that when coupled to the puncture
gauge and linearized around the Minkowski spacetime, the Z4
formulation is numerically stable with a particular discretization.
The calculations build on studies of numerical stability for first
order in time, second order in space
systems~\cite{CalHinHus05,ChiHus10,WitHilSpe10}, and essentially
rely on the theory of~\cite{KreLor89,GusKreOli95}.


\subsection{A novel discretization of second order systems
of tensors}\label{Section:Novel}


\paragraph*{Motivation:} For linear, constant coefficient
first order strongly hyperbolic systems, strong hyperbolicity
is enough to guarantee numerical stability under a
method of lines approach, which is no longer the case for first
order in time, second order in space systems. If in addition
to being strongly hyperbolic, the PDE system is symmetric
hyperbolic, then a discrete energy method can be used to
guarantee numerical stability. There are many second order
PDE systems of interest that are only strongly hyperbolic,
for which other methods must be used in analysis. Unfortunately
when using the standard discretization in space for strongly
hyperbolic second order systems it is often not even possible
to analyze whether or not the resulting semi-discrete system
is stable with computer algebra. The reason for this is that
there is not a straightforward relationship between the
principal symbol of the semi-discrete system and that of the
continuum; various sectors of the principal symbol that are
decoupled at the continuum become entangled in the
semi-discrete system. Here we present a novel discretization
for systems of tensors for which semi-discrete stability
analysis can at least be tackled. Intuitively this is made 
possible because after Fourier-transform the novel 
discretization has only one wave-vector, whereas the standard
discretization has two. The novel discretization
does not guarantee numerical stability whenever the continuum
system is strongly hyperbolic. Therefore the situation is
still not entirely satisfactory.

\paragraph*{The standard discretization:} For a
grid-function~$f$, where here and in the following we suppress indices
labeling the position on the grid, the standard second order
discretization accurate for second order in space systems is
\begin{align}
\label{eqn:standard_second}
\p_i\to D_{0i},\qquad \p_i\p_j \to
\left\{\begin{array}{cc}
  D_{0i}D_{0j}  &\text{if   } i\ne j, \\
   D_{+i}D_{-i} & \text{if   }i=j,
\end{array}
\right.
\end{align}
where we use the standard notation for centered and one-sided
finite difference operators. The standard fourth order accurate
discretization for second order in space
systems is
\begin{align}
\p_i&\to D^{(4)}_{i}\equiv D_{0i}
\left(1-\frac{h^2}{6}D_{+i}D_{-i}\right),\\
 \p_i\p_j &\to  D^{(4)}_{0i}\equiv
\left\{\begin{array}{cc}
 D^{(4)}_{0i}D^{(4)}_{0j}  &\text{if   } i\ne j, \\
 D_{+i}D_{-i}(1-\frac{h^2}{12}D_{+i}D_{-i}) & \text{if   }i=j,
\nonumber
\end{array}
\right.
\end{align}
where here we drop the summation convention. This approach to
discretization can of course be naturally extended to higher
order accuracy.

\paragraph*{A novel approach to the discretization of
tensors:} We are concerned with numerical approximations to the
spatial derivatives of scalars, vectors and symmetric two-tensors,
and assume that we have a metric~$\dot{\eta}_{ij}$ in space.
Suppose that we are given the finite difference
approximations~$D^{(1)}_{i}$ and $D^{(2)}_{ij}$ to the first and
second derivatives, respectively. For the first derivative we
must therefore always use~$D^{(1)}_{i}$. For the second derivatives
we distinguish as follows. The Laplace operator is approximated by
\begin{align}
\eta^{ij}\p_i\p_j&\to\Delta^{(2)} \equiv \eta^{ij}D^{(2)}_{ij}.
\end{align}
Second gradients of scalars are approximated by
\begin{align}
\p_i\p_ju&\to \big[D^{(1)}_iD^{(1)}_j\big]^{\text{tf}}u
+ \frac{1}{3}\eta_{ij}\Delta^{(2)}u.
\end{align}
For vectors we distinguish gradient-divergence terms by using
repeated application of the first derivative,
\begin{align}
\p_i\p_jf^j&\to D^{(1)}_iD^{(1)}_jf^j.
\end{align}
Likewise for symmetric two-tensors we approximate
divergence-divergence terms by repeated application of the
first derivative approximation, although such terms do not
appear in our applications. In~\cite{CalHinHus05} numerical
stability of a parametrized generalization of the KWB
formulation~\cite{KnaWalBau02} of electromagnetism was
discussed. It was found that with the standard discretization
there are choices of the formulation
parameter that render the system numerically unstable, despite
the initial value problem being well-posed at the continuum.
It is straightforward to show that with the novel
discretization, using the standard second order discretization
for the raw approximation, the semi-discrete system is
numerically stable for all choices of the parameters that
render the initial value problem well-posed.


\subsection{Application to the Z4 formulation}
\label{Section:Z4_analysis}


\paragraph*{The linearized system:} In this section we present
the Z4c field equations linearized around the line element
\begin{align}
\textrm{d}s^2&= -\textrm{d}t^2
+\dot{\eta}_{ij}\textrm{d}x^i\textrm{d}x^j,
\end{align}
where we denote the constant background spatial metric
by~$\dot{\eta}_{ij}$. The background shift, extrinsic curvature
and Z4 constraints are taken to vanish. We denote the perturbations
by~$\alpha,\beta^i,\gamma_{ij},\Theta,Z_i,K_{ij}$. The linearized
equations of motion are
\begin{align}
\p_t\gamma_{ij}&= -2 K_{ij} + 2\p_{(i}\beta_{j)},\\
\p_t\alpha&=-\mu_L(K-2\Theta),\\
\p_t\beta^i&= \bar{\mu}_S f^i,\\
\p_tK_{ij} &=-\frac{1}{2}\p^k\p_k\gamma_{ij}-\frac{1}{6}\p_i\p_j\gamma
-\p_i\p_j\alpha+\p_{(i}f_{j)},\\
\p_t\Theta &= -\frac{1}{3}\p^i\p_i\gamma+\frac{1}{2}\p_if^i,\\
\p_t f^i &= \p^j\p_j\beta^i + \frac{1}{3}\p_i\p_j\beta^j
-\frac{4}{3}\p^iK +2\p^i\Theta,
\end{align}
where here we have
defined~$\bar{\mu}_S=(\dot{\eta})^{\frac{1}{3}}\mu_S$, and  used the
Nagy-Ortiz-Reula (NOR)~\cite{NagOrtReu04} variable
\begin{align}
f^i &= 2Z^i+\dot{\eta}^{ij}\dot{\eta}^{kl}\big[\p_k\gamma_{lj}
-\frac{1}{3}\p_j\gamma_{kl}\big].
\end{align}
The linearized Hamiltonian and momentum constraints are
\begin{align}
H&=\p^i\p^j\gamma_{ij}-\p^i\p_i\gamma,\\
M_i&=\p^jK_{ij}-\p_iK.
\end{align}

\paragraph*{The semi-discrete system:} We now discretize the
system in space but leave time continuous. Using the novel
discretization described in section~\ref{Section:Novel} we write
the system as
\begin{align}
\p_t\gamma_{ij}&= -2 K_{ij} + 2D^{(1)}{}_{(i}\beta_{j)},\\
\p_t\alpha&=-\mu_L(K-2\Theta),\\
\p_t\beta^i&= \bar{\mu}_S f^i,\\
\p_tK_{ij}&=-\frac{1}{2}\Delta^{(2)}\gamma_{ij}
-\frac{1}{6}\big[[D^{(1)}{}_{i}D^{(1)}{}_{j}]^{\textrm{tf}}
+\frac{1}{3}\dot{\eta}_{ij}\Delta^{(2)}\big]\gamma\nonumber\\
&-\big[[D^{(1)}{}_{i}D^{(1)}{}_{j}]^{\textrm{tf}}
+\frac{1}{3}\dot{\eta}_{ij}\Delta^{(2)}\big]\alpha
+D^{(1)}{}{}_{(i}f_{j)},\\
\p_t\Theta &= -\frac{1}{3}\Delta^{(2)}\gamma
+\frac{1}{2}D^{(1)}{}_if^i,\\
\p_t f^i &= \Delta^{(2)}\beta^i
+\frac{1}{3}D^{(1)}{}^iD^{(1)}{}_j\beta^j
-\frac{4}{3}D^{(1)}{}^iK + 2D^{(1)}{}^i\Theta,
\end{align}
where we suppress indices labeling the grid, which we take
to be
\begin{align}
x_i&=(x_{i_1},y_{i_2},z_{i_3})=(i_1h,i_2h,i_3h),
\end{align}
with $i_i=0\dots N-1$, $h=\frac{2\pi}{N}$ the spatial
resolution, and restrict our attention to $2\pi$-periodic
solutions. Note that in contrast to the continuum system the
constraint subsystem does not close for the semi-discrete
system. It is not obvious how to achieve closure of the
constraint subsystem at the semi-discrete level without
using a ``$(D_0)^2$'' type discretization. Unfortunately,
as discussed in~\cite{CalHinHus05} the $D_0$ norm is not 
strong enough for analytic considerations, and therefore 
do not consider it further here. 

\paragraph*{Psuedo-discrete reduction to first order:} Performing
a discrete fourier transform in space, defining
\begin{align}
\Omega_+^2&=\sum_{i=1}^{3}|\tilde{D}_{+i}|^2,
\end{align}
and~$\xi_i=-\pi+2\pi/N,-\pi+4\pi/N,\dots,+\pi$, for $i=1,2,3$, and
then introducing the reduction variables
\begin{align}
\tilde{\epsilon}&=i\Omega_+\tilde{\alpha},\qquad
\tilde{\phi}^i=i\Omega_+\tilde{\beta}^i,\qquad
\tilde{\iota}_{ij}=i\Omega_+\tilde{\gamma}_{ij},
\end{align}
results in a large set of ODEs which can be written as
\begin{align}
\p_t\tilde{\gamma}_{ij}&=-2\tilde{K}_{ij}
+\frac{2s}{\Omega_+}\hat{s}_{(i}\tilde{\phi}_{j)},\\
\p_t\tilde{\alpha}&=-\mu_L(\tilde{K}-2\tilde{\Theta}),\\
\p_t\tilde{\beta}^{i}&=\bar{\mu}_S\tilde{f}^i,
\end{align}
for the metric components,
\begin{align}
\p_t\tilde{\iota}_{ij}&=-2i\Omega_+\tilde{K}_{ij}
+2is\hat{s}_{(i}\tilde{\phi}_{j)},\\
\p_t\tilde{\epsilon}&=
-i\Omega_+\mu_L(\tilde{K}-2\tilde{\Theta}),\\
\p_t\tilde{\phi}^{i}&=i\Omega_+\bar{\mu}_S\tilde{f}^i,
\end{align}
for the reduction variables, and finally
\begin{align}
\p_t\tilde{K}_{ij}&=
-\frac{i}{\Omega_+}\big[s^2[\hat{s}_i\hat{s}_j]^{\textrm{tf}}
+\frac{1}{3}\dot{\eta}_{ij}\Omega^2\big]\tilde{\epsilon}
-\frac{i}{2}\frac{\Omega^2}{\Omega_+}\tilde{\iota}_{ij}\nonumber\\
&-\frac{i}{6\Omega_+}\big[s^2[\hat{s}_i\hat{s}_j]^{\textrm{tf}}
+\frac{1}{3}\dot{\eta}_{ij}\Omega^2\big]\tilde{\iota}
+is\hat{s}_{(i}\tilde{f}_{j)},\\
\p_t\tilde{\Theta}&=-\frac{i}{3}\frac{\Omega^2}{\Omega_+}
\tilde{\iota}+\frac{1}{2}is\tilde{f}^{\hat{s}},\\
\p_t\tilde{f}^i&=i\frac{\Omega^2}{\Omega_+}\tilde{\phi}^i
+\frac{i}{3}\frac{s^2}{\Omega_+}\hat{s}^i\tilde{\phi}^{\hat{s}}
-\frac{4}{3}is\hat{s}^i\tilde{K}+2is\hat{s}^i\tilde{\Theta}.
\end{align}
for the remaining quantities. Here we have defined
\begin{align}
s_i= -i\tilde{D}^{(1)}_{i},\qquad \Omega^2= - \tilde{\Delta}^{(2)},
\qquad s^2 = s_i s^i,\qquad \hat{s}^i = \frac{s^i}{s},
\end{align}
and use an index~$\hat{s}$ to denote contraction with the unit
wave-vector $\hat{s}^i$. Whilst estimating the growth of the
solutions we may discard the non-principal terms because for the
standard second and fourth order accurate discretizations
the non-principal additions are bounded
(Theorem 5.1.2~\cite{GusKreOli95}). Note that for both the
standard second and fourth order accurate discretizations we have
the properties,
\begin{align}
\label{eqn:Disc_Lims}
0 \le \frac{s}{\Omega} \le a,\qquad
b^{-1} \le \frac{\Omega}{\Omega_+} \le b,
\end{align}
for some positive constants $a,b$, which we will use in the
following. Upto non-principal terms the evolution of the metric
components~$\tilde{\alpha},\tilde{\beta}^i,\tilde{\gamma }_{ij}$ is
trivial, so they may be discarded in the following discussion.
Defining the unit wave-vector~$\hat{s}_i$, and using it
with~$\dot{\eta}_{ij}$ as a metric in Fourier-space to perform
a~$2+1$ decomposition reveals that, under the novel discretization
the principal symbol of the pseudo-discrete reduction decouples into
a scalar
($\tilde{\iota}_{ss},\tilde{\iota}_{qq},\tilde{\epsilon},\tilde{f}^s,
\tilde{\hat{K}}_{ss},\tilde{\hat{K}}_{qq},
\tilde{\Theta},\tilde{\phi}^s$), vector
($\tilde{\iota}_{sA},\tilde{f}^A,\tilde{\hat{K}}_{sA},\tilde{\phi}^A$)
and tensor
($\tilde{\iota}^{\tiny{\text{TF}}}_{AB},\tilde{\hat{K}}^{\tiny{\text{TF}}}_{AB}$)
sector, just like in the continuum PDEs analysis.

\paragraph*{Characteristic variables:} Each of the three
sectors generically admits a complete set of characteristic
variables. In the scalar sector they are
\begin{align}
u_{\pm\sqrt{\mu_L}}&=\frac{\Omega}{\Omega_+}
\tilde{\epsilon}\pm\sqrt{\mu_L}
\tilde{\hat{K}},\label{eqn:lapse-mode}
\end{align}
which are associated with the lapse gauge condition, propagating with
speed~$\pm\sqrt{\mu_L}i\Omega$, and appear almost identically at the
continuum level. For brevity we use the
shorthands~$\tilde{\hat{K}}=\tilde{K}_{\hat{s}\hat{s}}+\tilde{K}_{qq}
-2\tilde{\Theta}$ and~$\tilde{\Lambda}=\tilde{\iota}_{\hat{s}\hat{s}}
+\tilde{\iota}_{qq}$. Next we find
\begin{align}
u_{\pm1}&=2\Omega\frac{3\Omega^2+s^2}{\Omega_+}\tilde{\iota}_{\hat{s}\hat{s}}
+2\Omega\frac{s^2-3\Omega^2}{\Omega_+}\tilde{\iota}_{qq}
\pm 12(\Omega^2-s^2)\tilde{K}_{\hat{s}\hat{s}}\nonumber\\
&\mp 6(\Omega^2+2s^2)\tilde{K}_{qq}\pm 24s^2\tilde{\Theta}
-12\Omega s\tilde{f}^{\hat{s}},
\end{align}
propagating with speeds~$\pm i\Omega$, which corresponds to light speed in
the continuum system. At the continuum four characteristic variables in
the scalar sector propagate at light speed. In the semi-discrete system
with our discrectization the third and fourth of these characteristics
\begin{align}
u_{\pm v_-}&=\frac{3\bar{\mu}_S\Omega^2-3v_-^2
+s^2(\bar{\mu}_S-1)}{6\Omega_+(v_-^2-\Omega^2)}v_-\tilde{\Lambda}\nonumber\\
&
+\frac{3\bar{\mu}_S\Omega^2-3v_-^2+s^2(\bar{\mu}_S-1)}{\Omega_+(v_-^2-\mu_L\Omega^2)}
v_-\tilde{\epsilon}
\pm\frac{s}{\Omega_+}\tilde{\phi}^{\hat{s}}\nonumber\\
\label{eqn:v_-mode}
&\pm\frac{4s^2(\bar{\mu}_S-1)}{(\mu_L-\bar{\mu}_S)(3v_-^2-4\Omega^2)
-s^2\mu_L(\bar{\mu}_S-1)}
\bar{\mu}_S\tilde{\hat{K}}\nonumber\\
&\pm 2\frac{3\bar{\mu}_S\Omega^2-3v_-^2
+\bar{\mu}_Ss^2}{\Omega^2-s^2}
\tilde{\Theta}+\frac{\bar{\mu}_S-1}{\Omega^2-v_-^2}v_-s\tilde{f}^{\hat{s}},
\end{align}
become coupled to the characteristic variables of the shift. They propagate
with speeds~$\pm v_-$, where we define $v_\pm$ by
\begin{align}
v_\pm^2&=\frac{1}{6}(4+3\bar{\mu}_S)\Omega^2+\frac{1}{6}(\bar{\mu}_S-1)s^2
\nonumber\\
&\pm\frac{1}{6}\sqrt{\big[(4+3\bar{\mu}_S)\Omega^2+(\bar{\mu}_S-1)s^2\big]^2
-48\bar{\mu}_S\Omega^4}.
\end{align}
Under the substitution $\Omega\to s$ the continuum result (in fourier
space) is recovered. The last of the characteristic variables in the
scalar sector are naturally
associated with the gamma driver condition. They are given by the
same expression as~$\eqref{eqn:v_-mode}$ under the
replacement~$v_-\to v_+$, and have speeds~$\pm v_+$. Special cases
in the choice of the gauge parameters that prevent the existence of
a complete set of characteristic variables in the scalar sector are
discussed below. Thankfully in the vector sector we have the
straightforward
\begin{align}
u_{A\,\pm 1}&=\frac{\Omega}{2\Omega_+}\tilde{\iota}_{\hat{s}A}
\pm\tilde{K}_{\hat{s}A}
-\frac{s}{2\Omega}\tilde{f}_A,\\
u_{A\,\pm \sqrt{\bar{\mu}_S}}&=\frac{\Omega}{\Omega_+}\tilde{\phi}_{A}
\pm\sqrt{\bar{\mu}_S}\tilde{f}_{A},
\end{align}
with speeds $\pm i\Omega$ and $\pm\sqrt{\bar{\mu}_S}i\Omega$. Finally in the
discretization of the gravitational wave degrees of freedom we have the
trivial,
\begin{align}
u^{\textrm{TF}}_{AB\,\pm 1}&=\frac{\Omega}{2\Omega_+}
\tilde{\iota}_{AB}^{\textrm{TF}}
\pm \tilde{K}_{AB}^{\textrm{TF}},
\end{align}
with speeds~$\pm i\Omega,$ and where TF denotes that the trace is
removed with respect to the projected background metric.

\begin{figure}[t]
\includegraphics[width=0.45\textwidth]{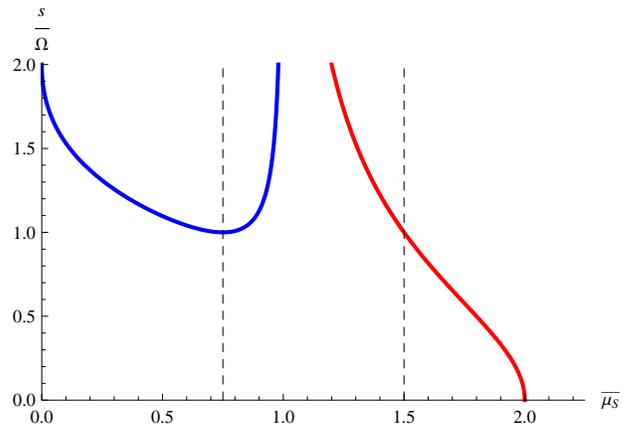}
\caption{\label{fig:Stab} The unstable gridmodes with the
novel discretization and~$\mu_L=2$. The left (blue) line
denotes~\eqref{eqn:unstab_modes_1} and the right
(red)~\eqref{eqn:unstab_modes_2}. The dashed vertical lines depict
the choices of~$\bar{\mu}_S$ that render the system weakly hyperbolic
at the continuum level. For those choices no stable discretization
is possible. In practice for the weakly hyperbolic choices of
$\bar{\mu}_S$ the characteristic variables become singular at
high resolution, preventing the construction of a symmetrizer
independent of~$h$. In principle one expects to be able to find
a stable discretization for all other values of $\bar{\mu}_S$,
but our novel discretization does not fulfil that requirement.
The unstable modes should be truncated in the $y$-axis at some
finite~$a\gtrsim 1$ determined by the discretization and background
metric, and so large values of unstable~$s/\Omega$ are not to be
considered troublesome. On the other hand behavior of the red line
in the region~$3\mu_L/4 <\bar{\mu}_S<\mu_L$ is undesirable, because
we always expect to find modes in that region on the numerical grid.
Note that for our standard choice, $\mu_S=1$, the blue
line diverges, provided that $\dot{\eta}\simeq 1$, implying that
our standard choice is stable, at least in that approximation.
There are never unstable modes when~$\mu_L<\bar{\mu}_S$.}
\end{figure}

\paragraph*{Special cases:} Assuming that $\mu_L>1$, there are
two special cases in the choice of the gauge
parameters~$\mu_L,\bar{\mu}_S$ under which the scalar sector does
not admit the complete set of characteristic
variables~(\ref{eqn:lapse-mode}-\ref{eqn:v_-mode})
given above. They are
\begin{align}
\label{eqn:unstab_modes_1}
\big[(4+3\bar{\mu}_S)\Omega^2+(\bar{\mu}_S-1)s^2\big]^2
-48\bar{\mu}_S\Omega^4
&=0,\\
\label{eqn:unstab_modes_2}
(3\mu_L-4)(\mu_L-\bar{\mu}_S)\Omega^2
-\mu_L(\bar{\mu}_S-1)s^2&=0,
\end{align}
which correspond
to~$4\bar{\mu}_S=3\alpha^2$ and~$3\mu_L=4\bar{\mu}_S$ at the
continuum in curved space. These special cases are more subtle in
the semi-discrete system because instability depends on both the
background metric and the specific order of the discretization.
Therefore we restrict our discussion here
to~$\dot{\eta}_{ij}=\delta_{ij}$, the gauge parameter $\mu_L=2$. Then
for either~$\bar{\mu}_S=\frac{3}{4}$ or~$\bar{\mu}_S=\frac{3}{2}$,
in the limit of high resolution, $h\to 0$ the characteristic variables
become degenerate in a neighborhood of the grid mode~$\xi_i=0$. This
situation is the same as that of the ADM formulation discussed
in~\cite{CalHinHus05}. Any other mode
satisfying~(\ref{eqn:unstab_modes_1}-\ref{eqn:unstab_modes_2})
will be unstable. There are no such modes for~$\mu_S<\frac{3}{4}$
or~$\frac{3}{4}<\mu_S<\frac{3}{2}$ because for the standard second
and fourth order discretizations~$a=1$ in~\eqref{eqn:Disc_Lims} above
with the Minkowski background. In the limit of high resolution one
also expects to find unstable modes in the region $\frac{3}{2}<\mu_S<2$,
although for particular choices of $\mu_S$ they may also appear at
low resolutions. The general case is sketched in
Fig.~\ref{fig:Stab}.

\paragraph*{Numerical Stability:} Notice that at the lowest and
highest grid-frequencies the principal symbol takes a different
form, because either $\Omega_+$ or $s^i$ vanish. As discussed
in~\cite{CalHinHus05} this poses no difficulty in the constant
coefficient case, provided that the principal symbol remains
diagonalizable for those frequencies, as the full solution
may be written as a direct sum over the modes. The principal
symbol vanishes at the lowest frequency and so is trivially
diagonalizable. At the highest frequency, by construction, the
principal symbol can be written as that of a system of decoupled
wave equations, and so is also trivially diagonalizable. The
argument given in~\cite{CalHinHus05} guarantees that standard
method of lines time-integrators will not violate the reduction
constraints. The existence of a complete set of characteristic
variables for the system for every grid frequency guarantees the
existence of a symmetrizer $\hat{H}(s_i)$, which in turn
guarantees numerical stability, namely that the estimate
\begin{align}
||u(t_n,\cdot)||_{h,D_+}&\leq K e^{\alpha t_n}||u(0,\cdot)||_{h,D_+},
\end{align}
holds for sufficiently small $h$, where $\alpha, K$ are constants,
independent of~$h$, that are not to be confused with the lapse and
extrinsic curvature, provided that the Courant factor~$\lambda$,
the ratio of the time-step and the spatial resolution, is chosen
sufficiently small. The Courant condition is~$\lambda\le
\alpha_0/(2\sqrt{3\mu_L})$, in the flat-space, second order
accurate case with~$\mu_S=1$ and~$\alpha_0$ the stability radius
of the time integrator. The norm~$||\cdot||_{h,D_+}$ is given by
\begin{align}
||u(t,\cdot)||^2_{h,D_+}&=||\alpha||^2_h+\sum_{i=1}^3||\beta^i||^2_h
+\sum_{i,j=1}^3||\gamma_{ij}||^2_h+||\Theta||^2_h\nonumber\\
&+\sum_{i=1}^3||f^i||^2_h
+\sum_{i,j=1}^3||K_{ij}||^2_h+\sum_{i=1}^3||D_{+i}\alpha||^2_h
\nonumber\\
&+\sum_{i,j=1}^3||D_{+i}\beta^j||^2_h
+\sum_{i,j,k=1}^3||D_{+i}\gamma_{jk}||^2_h.\label{eqn:D_+norm}
\end{align}

\paragraph*{Discussion:} The inclusion of a constant background
shift is straightforward since it affects the characteristic
speeds in a trivial way, and has no affect on the characteristic
variables. Different choices of discretization of the shift
advection derivatives were studied in detail in~\cite{ChiHus10}.
Setting the background lapse to unity was done only for convenience
and does not affect the results qualitatively. In the linear
constant coefficient regime the conformal variables can be expressed
as a linear combination of the NOR variables that we have used in our
calculations. The transformation to conformal variables is
{\it not} however just a change of variables, because the
transformation introduces the algebraic constraints~$D$ and~$T$.
In the linear constant coefficient case, as discussed
in~\cite{WitHilSpe10}, if an explicit polynomial time-integrator is
used then linearity of the system guarantees that the algebraic
constraints will not be violated if they are initially satisfied.
The time-integrator can furthermore be modified by a constraint
projection step, so that even if the initial data does include
violate the algebraic constraints then numerical stability of the
algorithm is trivially recovered. In the variable coefficient and
non-linear cases, the constraint projection step is essential
because error will otherwise violate the algebraic constraints, and
as shown in Appendix~\ref{App:weak}, when the algebraic constraints
are violated the Z4c formulation is generically only weakly
hyperbolic. It would be desirable to extend the results presented in
this section to the variable coefficient case, rather than
linearizing and freezing coefficients. The most powerful method for
doing so would be to use a summation by parts finite-differencing
scheme to conserve a semi-discrete energy. Unfortunately
such an approach does not work because, as shown in
Appendix~\ref{App:Not_Sym}, even at the continuum level no such
energy exists.


\section{The apples with apples tests}
\label{Section:AwA}



\subsection{Numerical setup}


For our numerical tests, we use the AMSS-NCKU
code~\cite{CaoYoYu08,GalBruCao10,CaoLiu11}. The code employs MPI
parallelization for moving box mesh refinement and method of lines
time integration. In physics applications the code uses the moving
puncture gauge with BSSNOK to evolve black hole spacetimes. Here it
is used in a much simpler context: each of the Apples with Apples
tests employs just a single mesh, and peridodic boundary conditions
in space, which avoids implementing complicated conditions in 3D.
Following~\cite{AlcAllBon03,BabHusAli08} we set simulation
domain~$x\in[-0.5,0.5]$. In contrast to the normal AMSS-NCKU setup,
we use an unstaggered, or vertex centered, grid~$x_n=-0.5+ndx$,
$n=0...50\rho$. We take~$dx=dy=dz=1/(50\rho)$, $\rho=1, 2, 4$, and
so on. In the~$y$ and~$z$ directions, following~\cite{BabHusAli08},
we use only five grid points. The Courant factor~$\lambda=dt/dx$ is
taken to be either~$0.1$, $0.25$, or~$0.5$ according to the specific
goal of each test. For spatial derivatives, we restrict exclusively
to second order accurate finite differencing. For advection
terms~$\beta^i\partial_i$, we use the upwind scheme
\begin{align}
D^{(up)}_i=\left\{\begin{array}{c}D_{+i}-\frac{1}{2}dx_iD_{+i}D_{+i}
\text{  if  }\beta^i>0,\\
D_{-i}+\frac{1}{2}dx_iD_{-i}D_{-i} \text{  if  } \beta^i<0.
\end{array}\right.
\end{align}
For the remaining spatial derivatives we employ either the standard
spatial discretization~\eqref{eqn:standard_second}, or the novel
scheme described in section~\ref{Section:Novel}. In~\cite{Hin05} it
was shown, albeit for gauge conditions other than those that we
employ here, that without artificial dissipation the BSSNOK
formulation is not numerically stable. Since we are comparing the
Z4c formulation with BSSNOK, we therefore always use Kreiss-Oliger
numerical dissipation~\cite{GusKreOli95}
\begin{align}
\label{eqn:KO_diss}
\partial_tu\rightarrow\partial_tu-\sigma\sum_{i=1}^{3}
dx_i^3(D_{i+}D_{i-})^2u,
\end{align}
with dissipation parameter~$\sigma=0.02$. In the original Apples
with Apples tests, gauge conditions were prescribed. Unfortunately
some of those prescriptions result in an ill-posed initial value
problem when coupled to either the BSSNOK or Z4c formulations.
Furthermore we prefer to study error properties of the exact system
that we will later use to perform astrophysical simulations. We
therefore use only the puncture
gauge~(\ref{eqn:BM_lapse}-\ref{eqn:Gamma_driver}) with
$\mu_L=2/\alpha$, $\mu_S=1/\alpha^2$ and~$\eta=2$ unless stated
otherwise. We are mainly concerned with the effect of each
formulation on the accuracy of the numerical calculation at finite
resolution and the effect of the novel discretization scheme versus
the standard discretization for the Z4c system. In our tests we
frequently use the constraint monitor
\begin{align}
\label{eqn:C_monitor}
C\equiv\left\{\begin{array}{cl}\sqrt{H^2+\gamma_{ij}M^iM^j+\Theta^2
+4\gamma_{ij}Z^iZ^j}& \text{  for Z4c,}\\
\sqrt{H^2+\gamma_{ij}M^iM^j+\gamma_{ij}G^iG^j}& \text{  for BSSNOK,}
\end{array}\right.
\end{align}
to assses the quality of the numerical solution. Here~$H$ is the
Hamiltonian constraint violation, $M^i$ is the momentum constraint
violation and $G^i$ is the Gamma constraint violation. Since~$2Z^i$
corresponds to the Gamma constraint in BSSNOK formalism, we use
$4\gamma_{ij}Z^iZ^j$ to mimic the Gamma constraint violation part.


\subsection{Robust stability test}


\begin{figure*}[t]
\begin{tabular}{cc}
\includegraphics[width=0.5\textwidth]{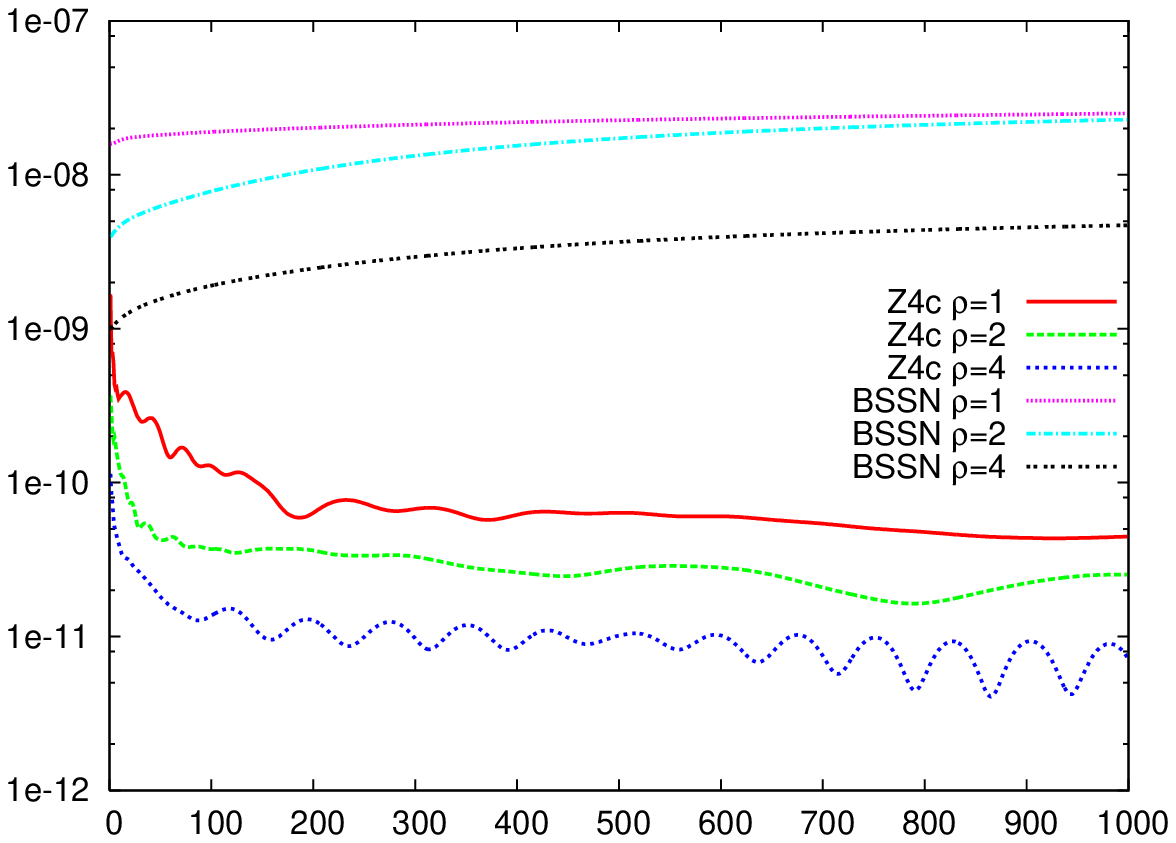}&
\includegraphics[width=0.5\textwidth]{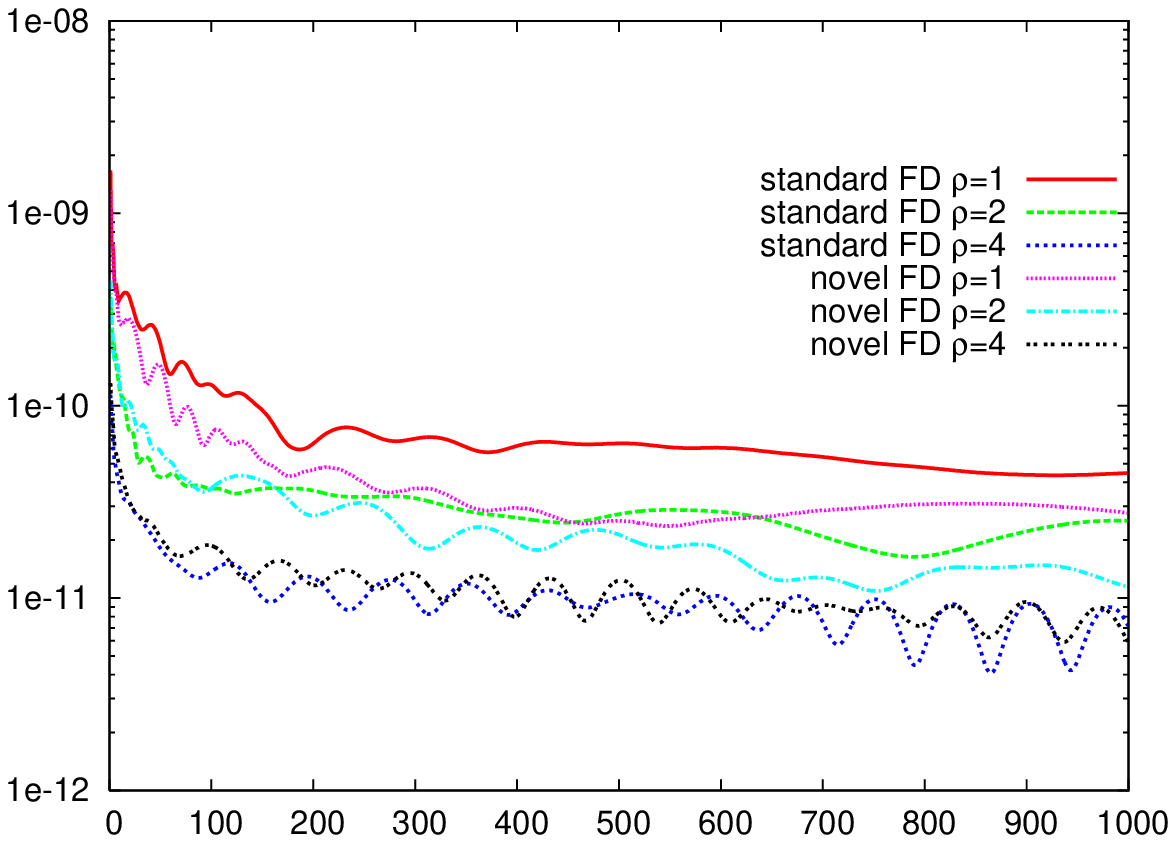}
\end{tabular}
\caption{ \label{fig:Robust} Comparison of the constraint violation
monitor~\eqref{eqn:C_monitor} for the robust stability test
with~$\eta=2$. The left subplot compares BSSNOK and Z4c, where the
standard discretization scheme is used. Numerical experiments reveal
that the drop in the constraint monitor in the Z4c tests is caused
neither by artificial dissipation nor the~$\eta$ parameter. The
right subplot compares the standard discretization scheme and the
novel discretization scheme for Z4c formalism.}
\end{figure*}

First we perform the robust stability test. Besides the gauge choice
we follow the original presciption. Initially all variables
including lapse and shift are set to the Minkowski value perturbed
by a random number $\epsilon\in(-10^{-10}/\rho^2,10^{-10}/\rho^2)$.
We set the Courant factor to $\lambda=0.5$. In
Fig.~\ref{fig:Robust}, the constraint violation monitor respect to
time is plotted for evolutions with~$\eta=2$. In the left subplot we
compare Z4c formalism with BSSNOK formalism using in each case the
standard discretization scheme. Although neither BSSNOK nor Z4c
suffer from rapid undesirable growth of the solution, the constraint
violations are slowly increasing for BSSNOK formalism while they are
decreasing for the Z4c test. By the end of the evolution, the
constraint violation in the Z4c tests are several orders of
magnitude lower than those of the BSSNOK test. We do not have an
explanation for this behavior. The usual argument is that the
propagating constraint subsystem of the Z4c formulation should
prevent constraint violation from growing on the grid, but in this
test the constraint violation can not be absorbed by the boundary
conditions. To investigate the cause of the decay of the constraints
we have performed tests without artificial dissipation. There we
find that the constraint violation levels off around an order of
magnitude above the value obtained with artificial dissipation. We
also tried evolutions with~$\eta=0$, and find that the constraint
monitor is larger than with~$\eta=2$, but that the qualitative
behavior is unaltered. Experimenting with the constraint damping
scheme with~$\kappa_1=0.02,\kappa_2=0$, we find a minor improvement
in the constraint monitor, in line with our expectation from the
results of~\cite{WeyBerHil11}. In the right subplot, we compare the
standard discretization scheme and the novel scheme, for which we
have a proof of numerical stability in the constant coefficient
approximation, each with the Z4c formalism. Here however we would
like to reiterate the important point made in~\cite{CalHinHus05},
namely that formal numerical stability neither implies, nor is
implied by robust stability. The robust stability test, in the form
studied here, should be viewed only as a gauge of the errors present
in the numerical evolution at a finite resolution. In
appendix~\ref{App:weak} we present more demanding convergence tests.


\subsection{Linear wave test}


\begin{figure}[t]
\begin{tabular}{cc}
\includegraphics[width=0.45\textwidth]{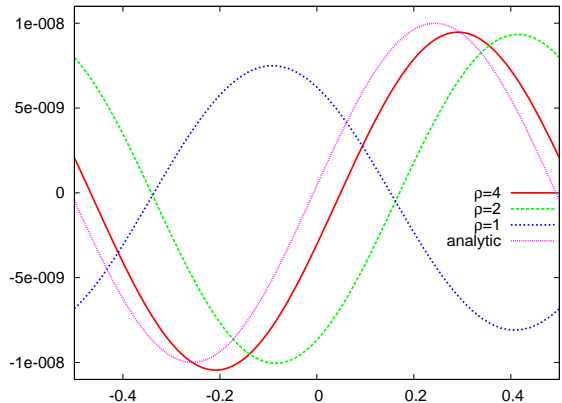}
\end{tabular}
\caption{
\label{fig:GW_profile}
Comparison snapshots of $\gamma_{yy}-1$ at $t=1000$
for Z4c formalism with the standard discretization scheme.}
\end{figure}


\begin{figure*}[ht]
\begin{tabular}{cc}
\includegraphics[width=0.5\textwidth]{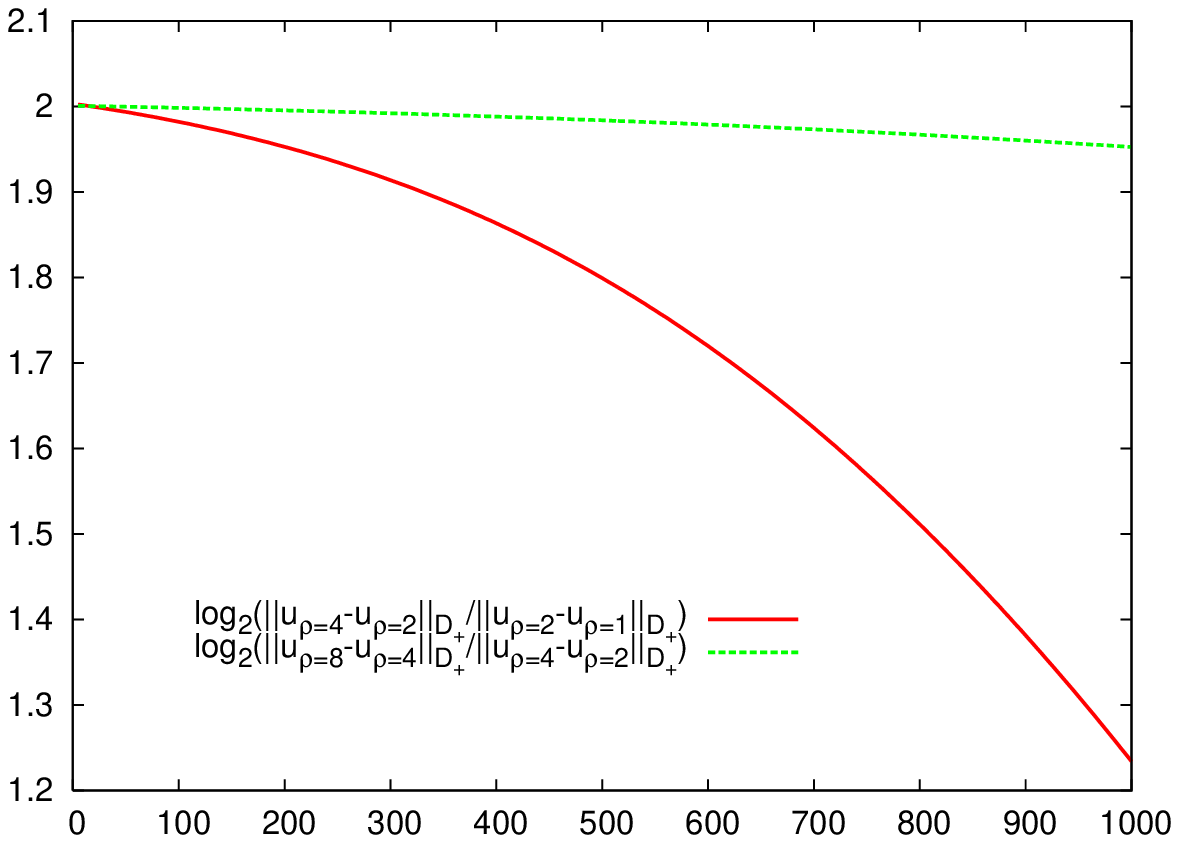}&
\includegraphics[width=0.5\textwidth]{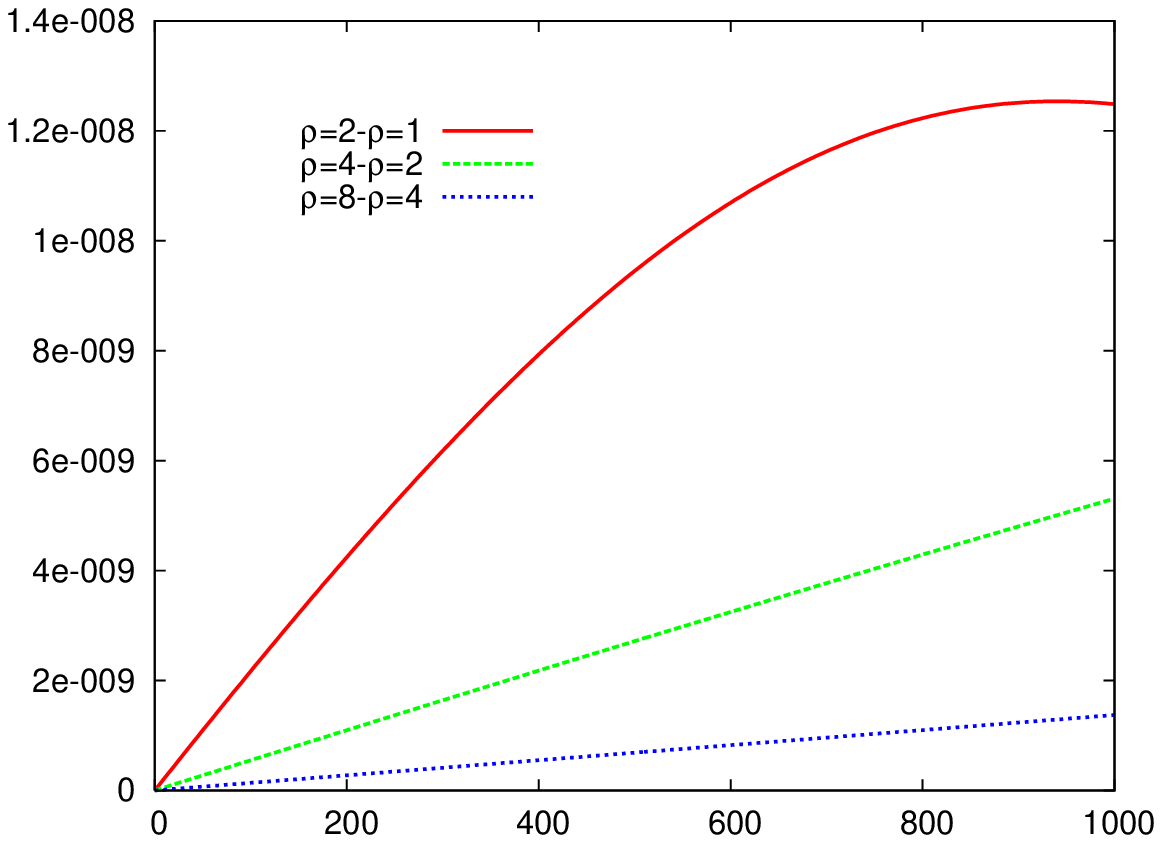}
\end{tabular}
\caption{x direction sine linear wave test with amplitude
$A=10^{-8}$. Puncture gauge is used. $\rho=1,2,4,8$ respectively.
Z4c formalism with standard discretization is used.}\label{fig2}
\end{figure*}

The second test is the evolution of a linearized gravitational wave
on the Minkowski background. We set initial data as
\begin{align}
&\gamma_{xx}=1,&\qquad
&\gamma_{yy}=1+b, &\qquad
&\gamma_{zz}=1-b,\nonumber\\
&\alpha=1,&\qquad
&K_{yy}=\frac{1}{2}\partial_t b,
&\qquad
&K_{zz}=-\frac{1}{2}\partial_t b,
\end{align}
and the remaining variables to zero. Here we choose
\begin{align}
\label{eqn:pert_expr}
b=A\sin\bigg[\frac{2\pi(x-t)}{d}\bigg],
\end{align}
and $d=1$, $A=10^{-8}$. The Courant factor is chosen~$\lambda=0.5$.
Such a small perturbation of the metric ensures that the numerical
evolution remains essentially in the linear regime. Here the initial
data are constraint violating, and the puncture gauge condition is
not necessarily compatible with simple advection of the wave
profile. In practice however it appears that to a good approximation
the solution is a simple travelling wave; in
Fig.~\ref{fig:GW_profile} we compare the final~$\gamma_{yy}$
numerical profile, obtained with Z4c, to the analytic solution of
the linearized Einstein equations with unit lapse and zero shift.
There it appears that at least the phase of the numerical solution
is converging with resolution to the analytic solution. A more
meaningful evaluation of the errors in the system is obtained by
performing self-convergence tests on the data, for which we use the
$D_+$ norm~\eqref{eqn:D_+norm} evaluated on the conformal variables
of the Z4c and BSSNOK formulations. There we find perfect second
order convergence in the norms for both BSSNOK and Z4c
(Fig.~\ref{fig2}, BSSNOK gives the same behavior as Z4c, not shown),
and estimate that after~$1000$ light crossing times, the error is
approximately $2\times10^{-9}$ for resolution $\rho=4$ and $8$
(Fig.~\ref{fig2}). Since the initial data for these evolutions is
constraint violating we do not present the constraint monitor for
this test, although we find that here it stays
around~$1\times10^{-12}$ with either formulation. The novel
discretization for Z4c gives almost identical results to those of
the standard discretization. We tested the effect of constraint
damping terms and different values of~$\eta$ in the gamma driver
shift. Neither has an effect on the result of these tests.
Concerning coordinate gauge conditions, for comparison, we have
performed additional evolutions with harmonic lapse, $\mu_L=1$
in~\eqref{eqn:BM_lapse} with harmonic shift~\eqref{eqn:Harm_Shift}.
We find both choices give very similar same convergence behavior and
errors.

%
\begin{figure*}[ht]
\begin{tabular}{cc}
\includegraphics[width=0.5\textwidth]{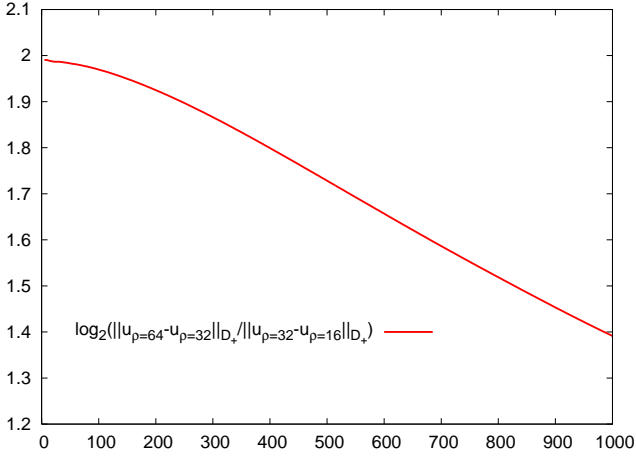}&
\includegraphics[width=0.5\textwidth]{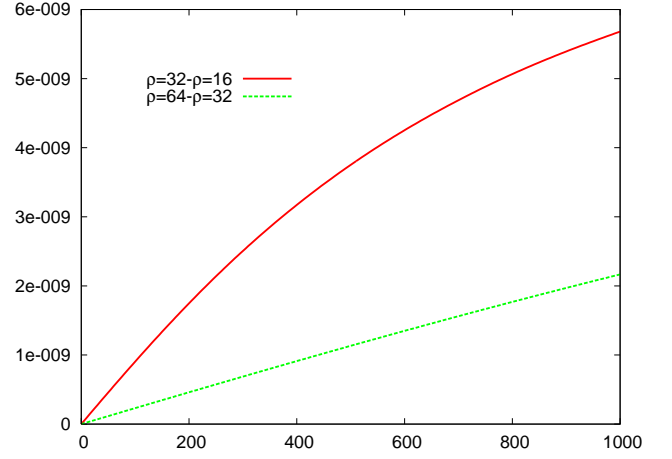}
\end{tabular}
\caption{Convergence factor of gauss shaped linear wave test with
amplitude $A=10^{-8}$. Puncture gauge is used.}\label{fig3}
\end{figure*}

In addition to above sine wave test, we also adopted the Gaussian
shaped linear wave test suggested in~\cite{BoyLinPfe06}. Instead of
Eq.(\ref{eqn:pert_expr}), the $b$ takes
\begin{align}
\label{eqn:gauss_expr} b=A\exp[-\frac{(x-t)^2}{2w^2}],
\end{align}
with $A=10^{-8}$. In order to get periodic profile initially we set
$w=0.5$ which makes $b$ roughly 0 at $x=\pm0.5$. Since the effective
wavelength for this gaussian wave is 1 which is one tenth of the
wave length of above sine wave, we need larger resolution to get
reasonably convergent result. Reducing numerical dispersion requires
more resolution still. In all we find resolutions $\rho=16$, $32$ 
and~$64$ can
produce good convergent result (Fig.~\ref{fig3}). The error for
$\rho=64$ and $32$ is comparable to the $\rho=8$ and $4$ of sine
wave case, which is roughly $2\times10^{-9}$.

%
\begin{figure*}[ht]
\begin{tabular}{cc}
\includegraphics[width=0.5\textwidth]{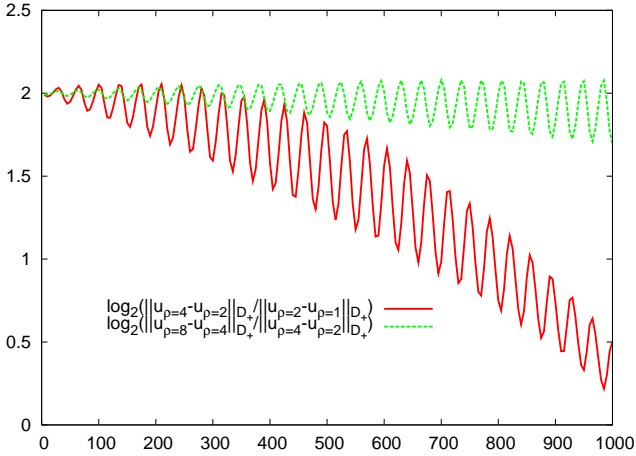}&
\includegraphics[width=0.5\textwidth]{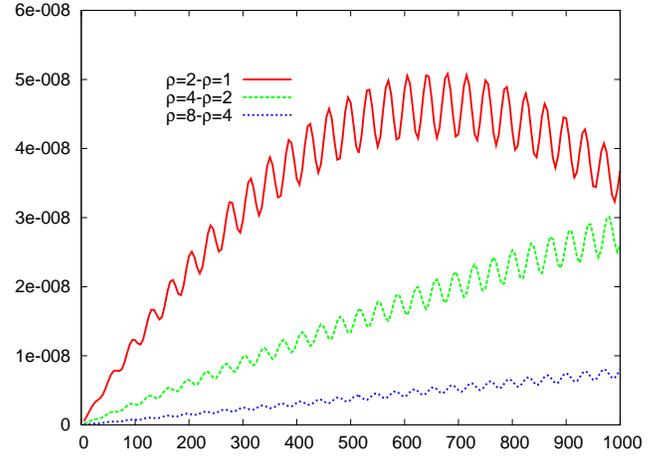}
\end{tabular}
\caption{Convergence factors and and errors for as 
in Fig.~\ref{fig3} but for the diagonal sine wave test.}\label{fig6}
\end{figure*}

As to the two dimensional sine wave test
\begin{align}
&\gamma_{xx}=\gamma_{yy}=1+\frac{H}{2}, \gamma_{xy}=\frac{H}{2},
& \gamma_{zz}=1-H, \nonumber\\
&K_{xx}=K_{yy}=K_{xy}=\frac{1}{4}\partial_t H,&
K_{zz}=-\frac{1}{2}\partial_t H,\\
& H=A\sin\bigg[\frac{2\pi(x-y-\sqrt{2}t)}{d}\bigg],
&\alpha=1,\nonumber
\end{align}
$d=1$, $A=10^{-8}$ and other variables zero, the result is similar
to the one dimensional test. But because the effective wave length
for two dimensional sine wave is $1/\sqrt{2}$, a little higher
resolution ($\rho=2,4,8$) is needed to get comparable convergence
behavior as in the one dimensional case (Fig.~\ref{fig6}).
As in the earlier robust stability tests the constraint damping 
scheme with~$\kappa_1=0.02,\kappa_2=0$ reduces the constraint 
violation very little.


\subsection{Gauge wave, and shifted gauge wave tests}



\begin{figure*}[ht]
\begin{tabular}{cc}
\includegraphics[width=0.5\textwidth]{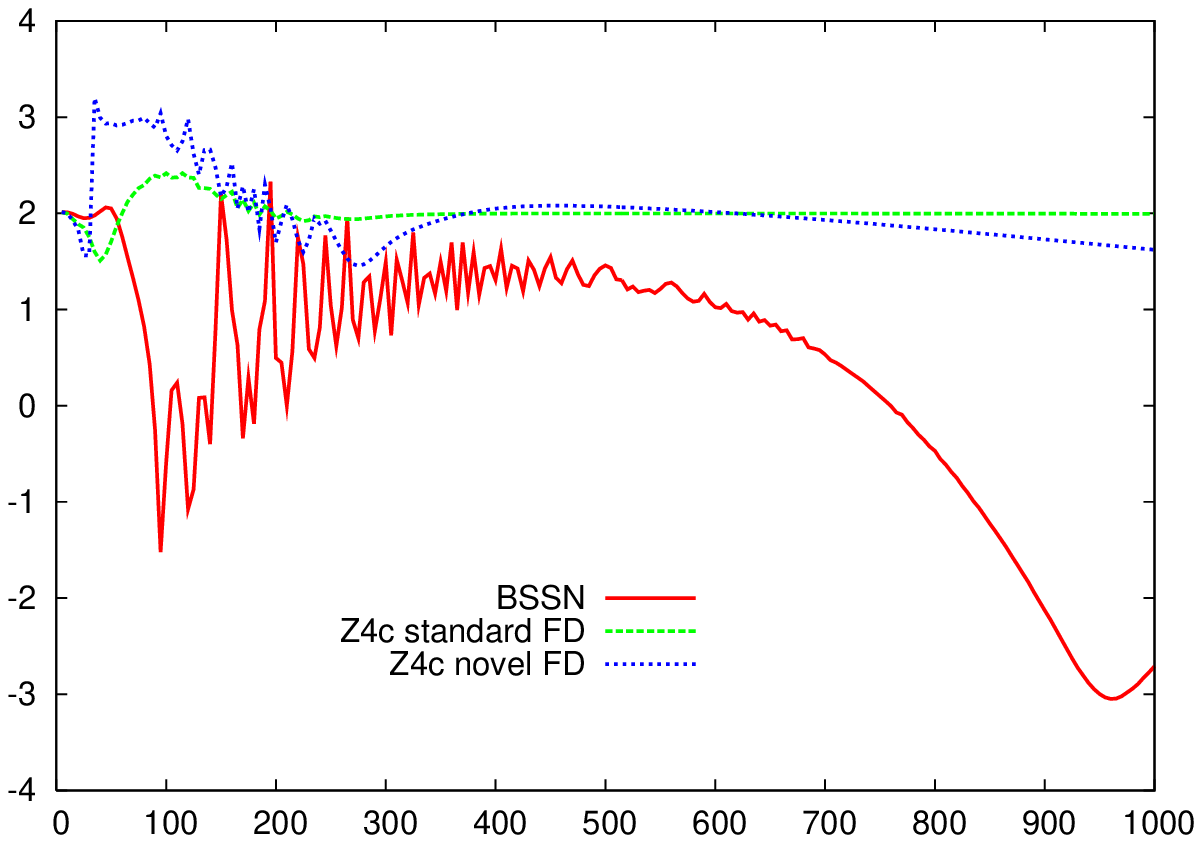}&
\includegraphics[width=0.5\textwidth]{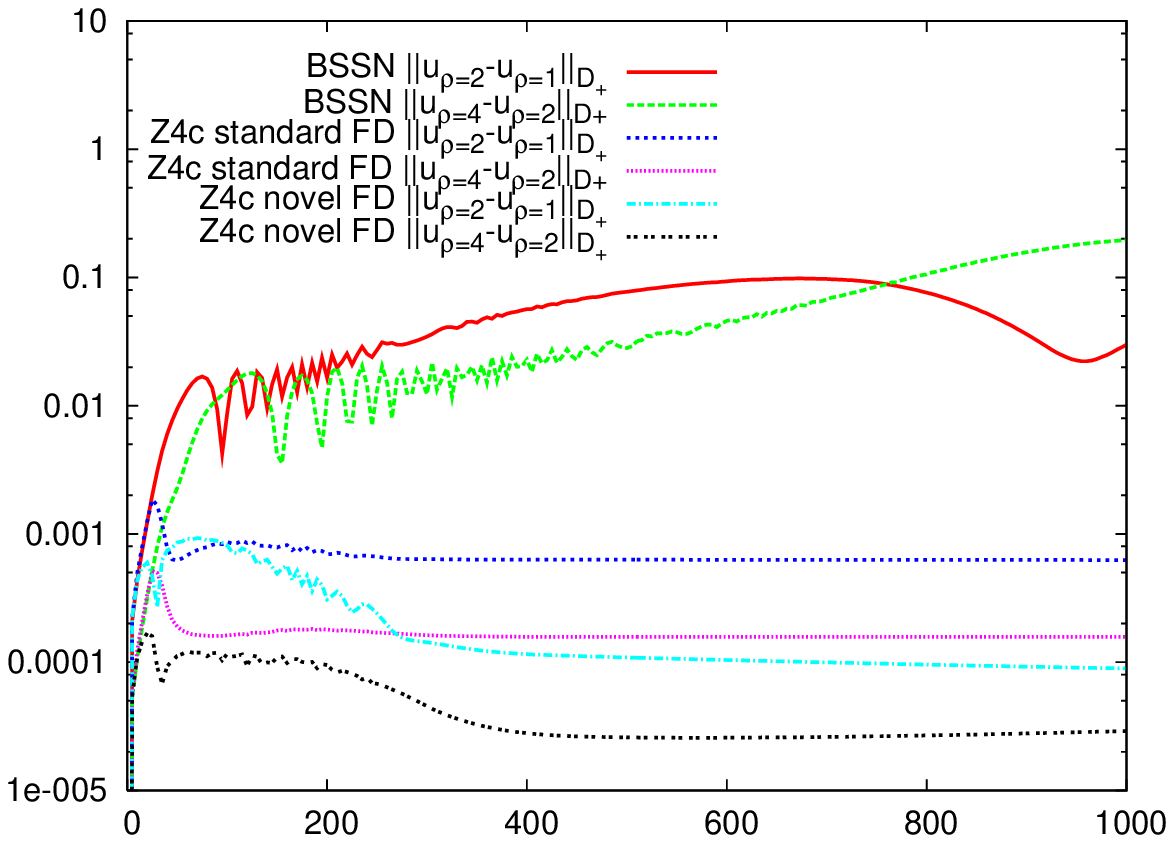}
\end{tabular}
\caption{One dimensional gauge wave test with $A=0.01$. Puncture
gauge is used. Left subplot: convergence factor of
$\log_2(\frac{||u_{\rho=4}-u_{\rho=2}||_{D_+}}{||u_{\rho=2}-u_{\rho=1}||_{D+}})$.
right subplot: $D_+$ norm of $u_{\rho=2}-u_{\rho=1}$ and
$u_{\rho=4}-u_{\rho=2}$.}\label{fig7}
\end{figure*}

\begin{figure*}[ht]
\begin{tabular}{cc}
\includegraphics[width=0.5\textwidth]{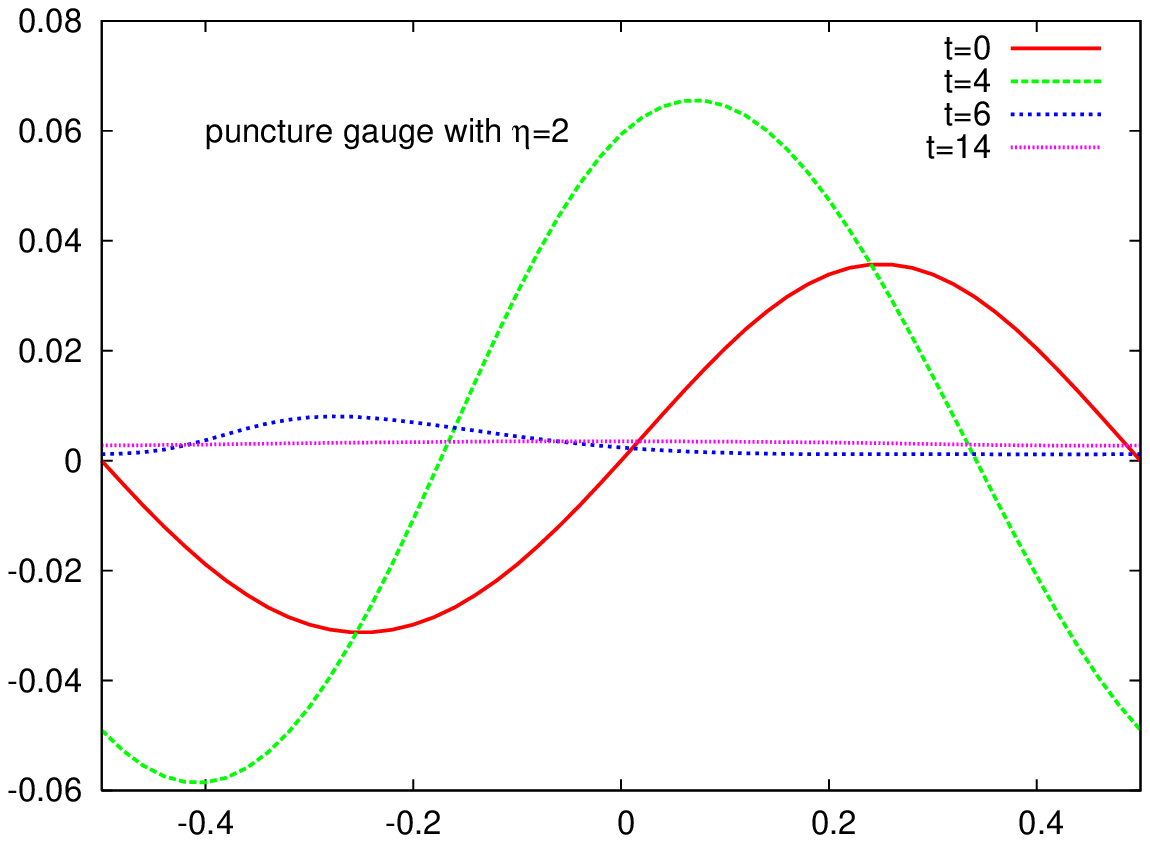}&
\includegraphics[width=0.5\textwidth]{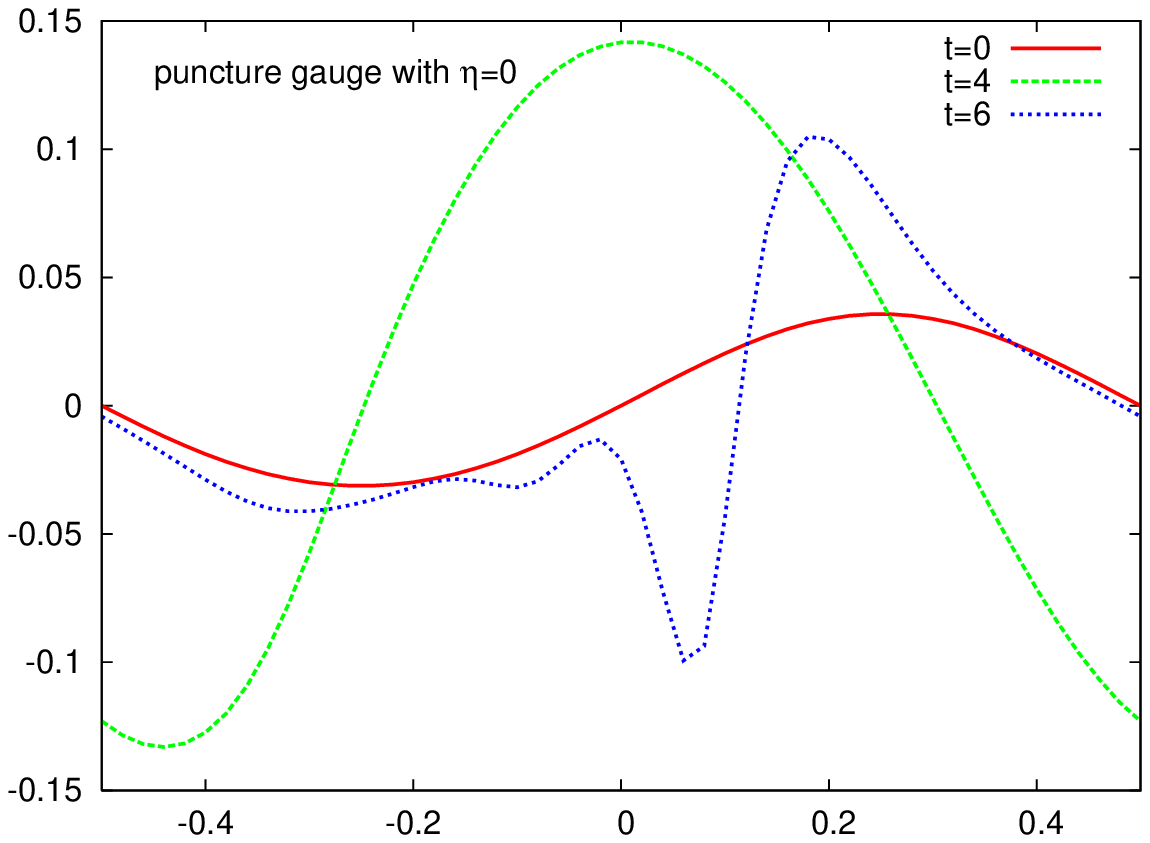}\\
\includegraphics[width=0.5\textwidth]{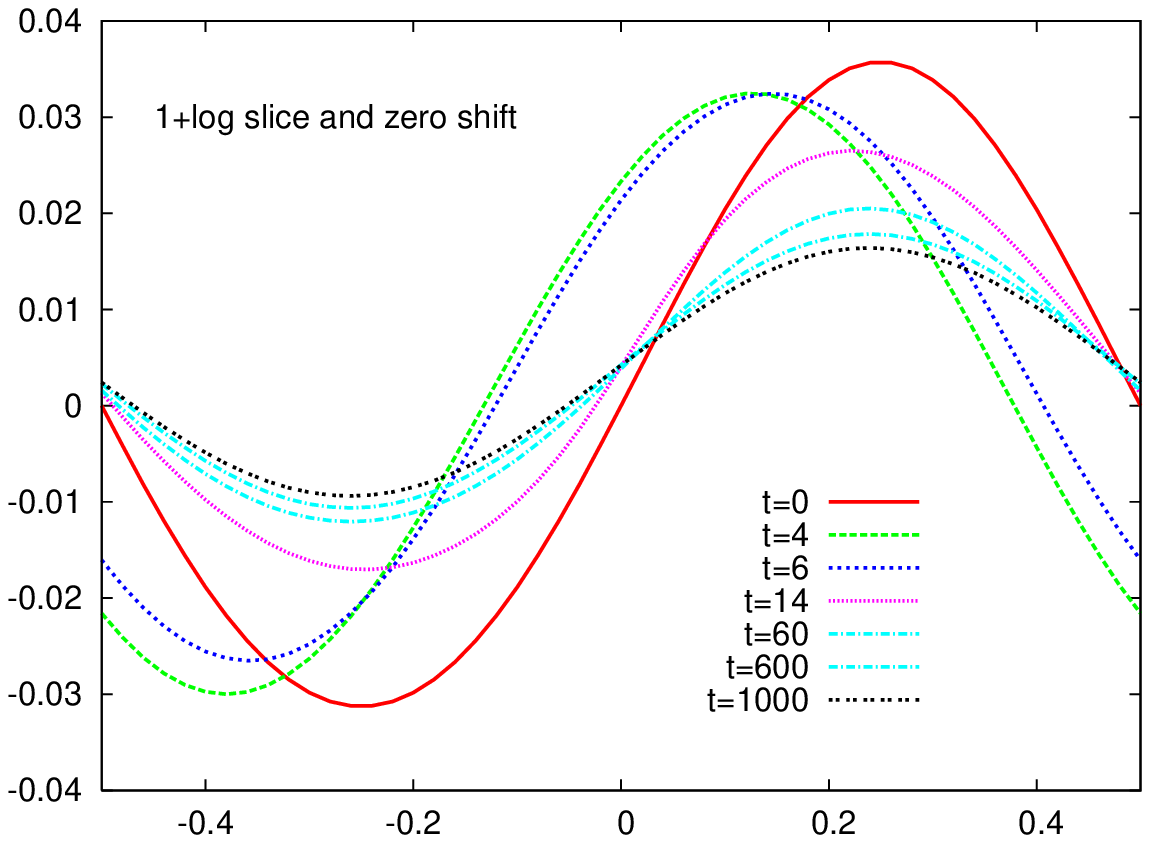}&
\includegraphics[width=0.5\textwidth]{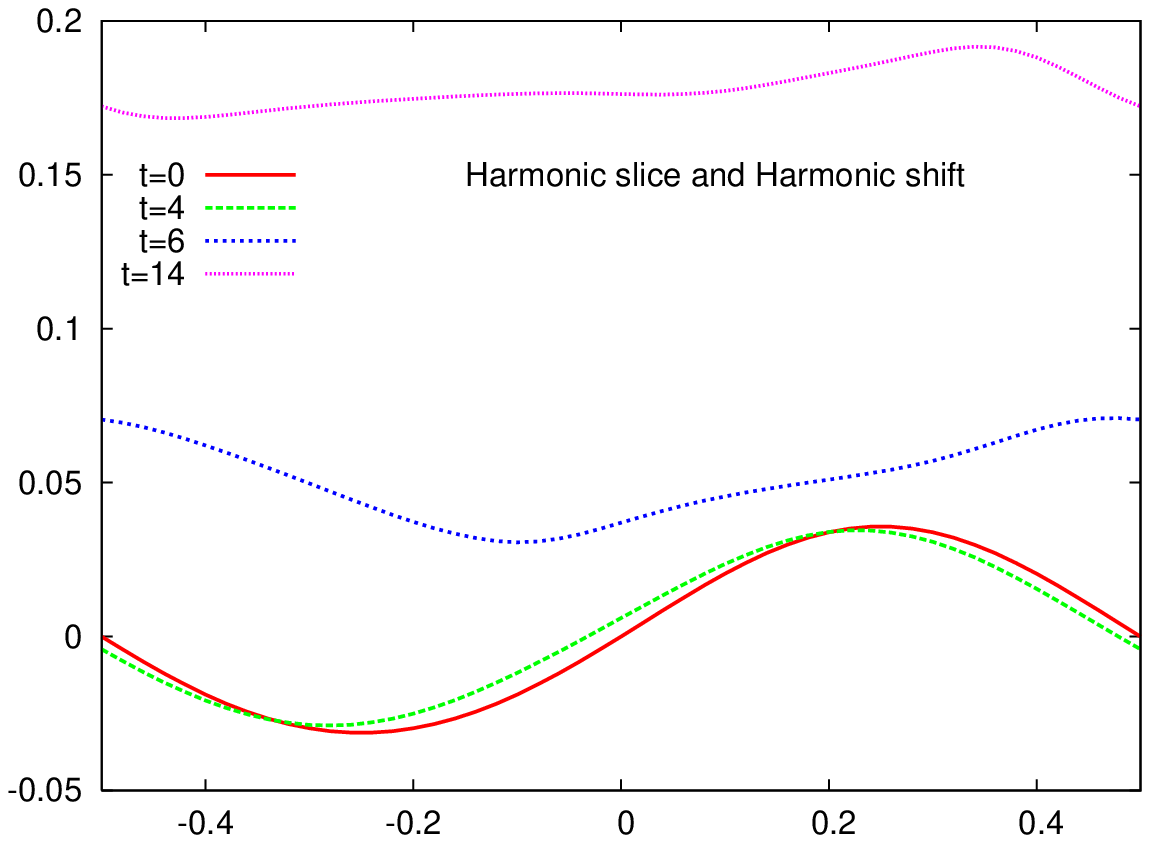}
\end{tabular}
\caption{\label{fig:Gauge_inertial} Gauge wave test with the
puncture gauge condition~(\ref{eqn:BM_lapse}-\ref{eqn:Gamma_driver},
$\eta=2$ or 0), with 1+log slice and zero shift and with harmonic
lapse and harmonic shift. Here~$\gamma_{yy}-1$ is plotted at times
$0$, $2$, $6$ and~$14$ respectively, for a test performed with the
Z4c formalism and the standard discretization scheme. With puncture
gauge and $\eta=2$ the gauge wave is rapidly suppressed by the gauge
condition and at later times $\gamma_{yy}-1$ tends to zero. With
$\eta=0$ the code crashes soon after $t=7.2$. While with other gauge
conditions we can not see this symmetry seeking property.}
\end{figure*}

The third and fourth tests are the unshifted gauge wave and
shifted gauge wave tests. For the gauge wave test we take
\begin{align}
&\gamma_{xx}=1-b, &\quad
&\gamma_{yy}=1,   &\quad
&\gamma_{zz}=1,\nonumber\\
&K_{xx}=\frac{\partial_t b}{2\sqrt{1-b}},&\quad  &\alpha=\sqrt{1-b},
&\quad & & &
\end{align}
as initial data. For the shifted gauge wave we use
\begin{align}
&\gamma_{xx}=1+b, &\quad
&\gamma_{yy}=1,    &\quad
&\gamma_{zz}=1,\\
&K_{xx}=\frac{\p_tb}{2\sqrt{1+b}}, &\quad
&\alpha=\frac{1}{\sqrt{1+b}}, &\quad
&\beta^x=-\frac{b}{1+b},\nonumber
\end{align}
and in either case set the the remaining variables to zero. We again
choose~\eqref{eqn:pert_expr}, but now with~$d=1$. Both~$A=0.01$ and
$A=0.1$ are tested. We set the Courant factor~$\lambda=0.5$ in the
unshifted case and~$\lambda=0.25$ in the evolution with non-trivial
initial shift. Otherwise the grid setup is the same as in the robust
stability test. It is well-known that with Harmonic lapse,
$\mu_L=1$, and vanishing shift the BSSNOK formulation suffers from
large undesirable growth of the errors in both of these
tests~\cite{BabHusAli08}. We recover that behavior for BSSNOK, and
find that with the Z4c system the blow-up is delayed. For a direct 
comparison with the gauge wave results of~\cite{AliBonBon11} we 
performed this test with Harmonic lapse, vanishing shift and constraint 
damping and~$\kappa_1=1,\kappa_2=0$. There, in agreement 
with~\cite{AliBonBon11} we find that the errors do not result in a 
code crash, although at this resolution the experimental convergence
factor is very poor. In contrast to the Harmonic lapse gauge condition, 
the puncture gauge is able to evolve the gauge wave for~$1000$ light 
crossing times with no sign of blow-up without constraint damping. 
Since both $A=0.01$ and $A=0.1$ give roughly the same result, in 
Fig.~\ref{fig7} we plot the convergence factor and the
$D_+$ norm of the error only for $A=0.01$. As the first test of 
nonlinear effects in ``apples with apples" test suite, Z4c gives 
better convergence behavior than BSSNOK. As shown in the left 
subplot of Fig.~\ref{fig7}, Z4c can approximately maintains a 
convergence factor around two while BSSNOK drops to negative values. 
Moreover, Z4c gives more than one order smaller error as shown in the 
right subplot of Fig.~\ref{fig7}. In terms of constraint violation we 
find that without Z4c constraint damping the constraint monitor is
comparable for the two formulations throughout the evolutions. On
the other hand we find that with $\kappa_1=0.02,\kappa_2=0.0$, the
Z4c constraint violation is an order of magnitude smaller than that
of BSSNOK at the end of the evolution. The experimental convergence
factor is unaffected by the use of the constraint damping scheme 
in this test. Furthermore we find that with the puncture gauge 
condition the coordinates are rapidly driven to match inertial 
coordinates of Minkowski space. In others words, the amplitude of 
the gauge wave decreases, as shown in Fig.~\ref{fig:Gauge_inertial}, 
so in some sense, the puncture gauge condition possesses a symmetry 
seeking property. As shown in this figure, we find that other gauge
conditions do not have this property and the nonzero value of~$\eta$
is also essential for the successful evolution. Note that this gauge
condition is similar to that used in binary black hole simulations.
The novel discretization gives near identical results to the
standard discretization. Tests with the harmonic gauge result in
both larger absolute error and larger constraint violations.

As for the two dimensional linear wave test, we use
\begin{align}
&\gamma_{xx}=\gamma_{yy}=1-\frac{H}{2}, \gamma_{xy}=\frac{H}{2},
& \gamma_{zz}=1, \nonumber\\
&K_{xx}=K_{yy}=-K_{xy}=\frac{\p_tH}{4\sqrt{1-H}},&
K_{zz}=-\frac{1}{2}\partial_t H,\\
& H=A\sin\bigg[\frac{2\pi(x-y-\sqrt{2}t)}{d}\bigg],
&\alpha=\sqrt{1-H},\nonumber
\end{align}
with $d=1$, $A=0.01,0.1$ for initial data for the two dimensional
gauge wave test, which corresponds to the wave propagating along
the diagonal direction of $x-y$ plane. The results are similar
to the one dimensional tests.

\begin{figure}[ht]
\begin{tabular}{c}
\includegraphics[width=0.45\textwidth]{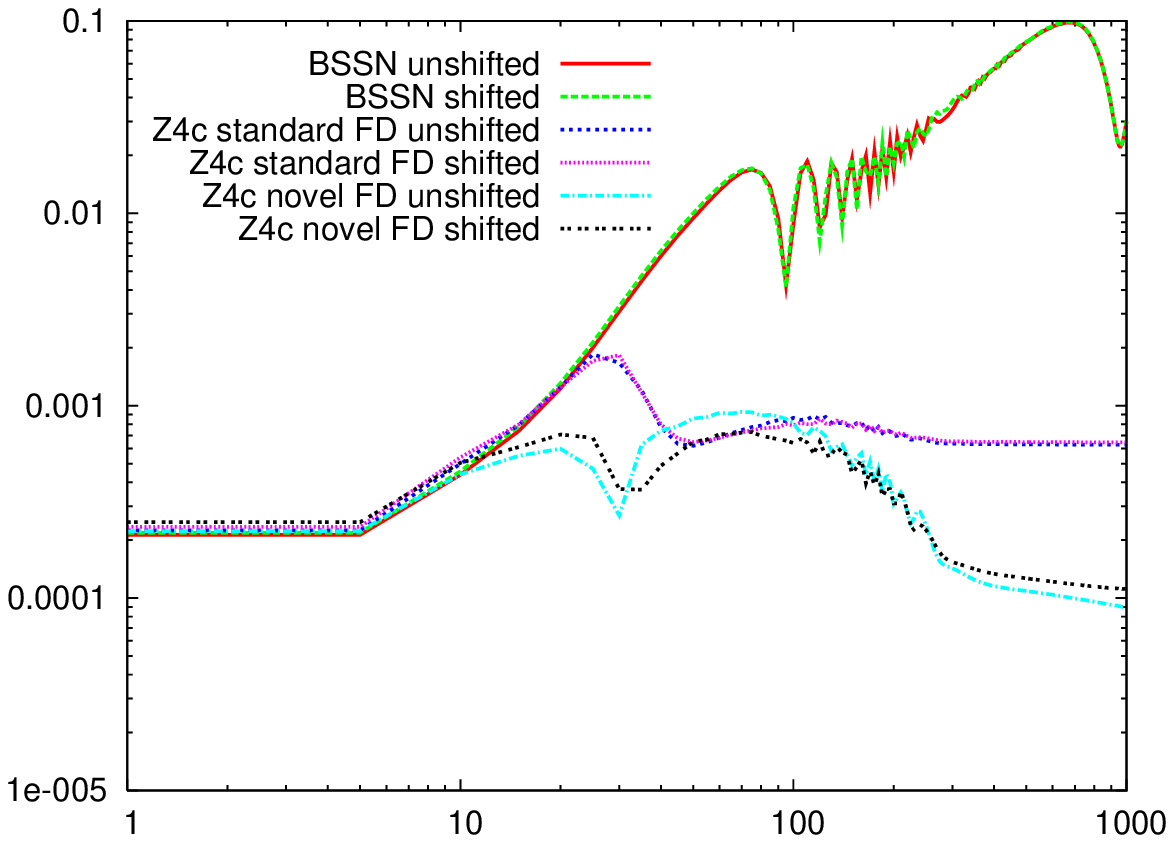}
\end{tabular}
\caption{ \label{fig8} Comparison of the $D_+$ norm of the error
for $u_{\rho=2}-u_{\rho=1}$ between unshifted gauge wave test
and shifted gauge wave test.}
\end{figure}

As for the shifted gauge wave test, original ``apples with apples"
test suggested~$A=0.5$. We find such a large amplitude results
in code crashes for both formalisms with any coordinate gauge.
Our findings are similar to those found with a pseudo-spectral
method as used in the SpEC code~\cite{BoyLinPfe06}. The unshifted gauge
wave tests also fail with amplitude~$A=0.5$. Restricting
to~$A=0.1,0.01$, we find the resulting error and the convergence
behavior is very similar to unshifted case. As an example we
compare the two results with $A=0.01$ in Fig.~\ref{fig8}.

%


\subsection{Gowdy wave test}


\begin{figure*}[t]
\begin{tabular}{cc}
\includegraphics[width=0.5\textwidth]{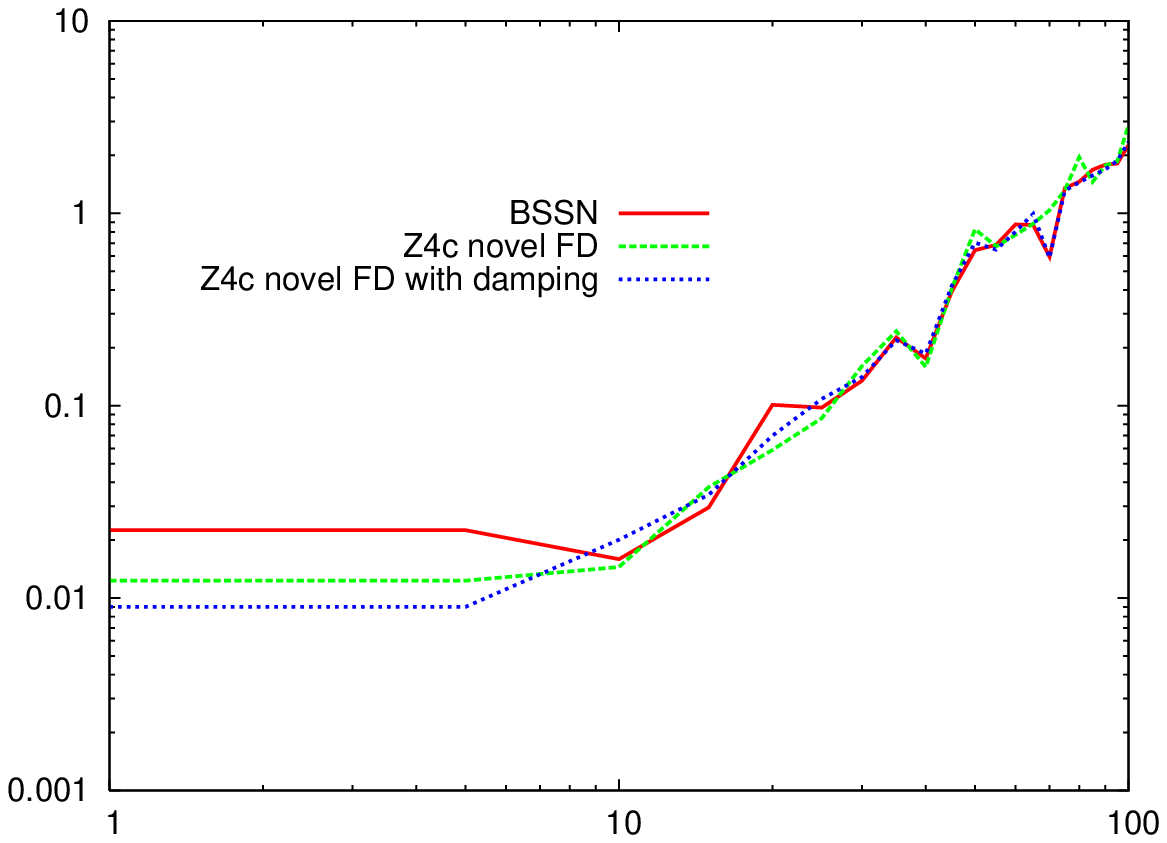}&
\includegraphics[width=0.5\textwidth]{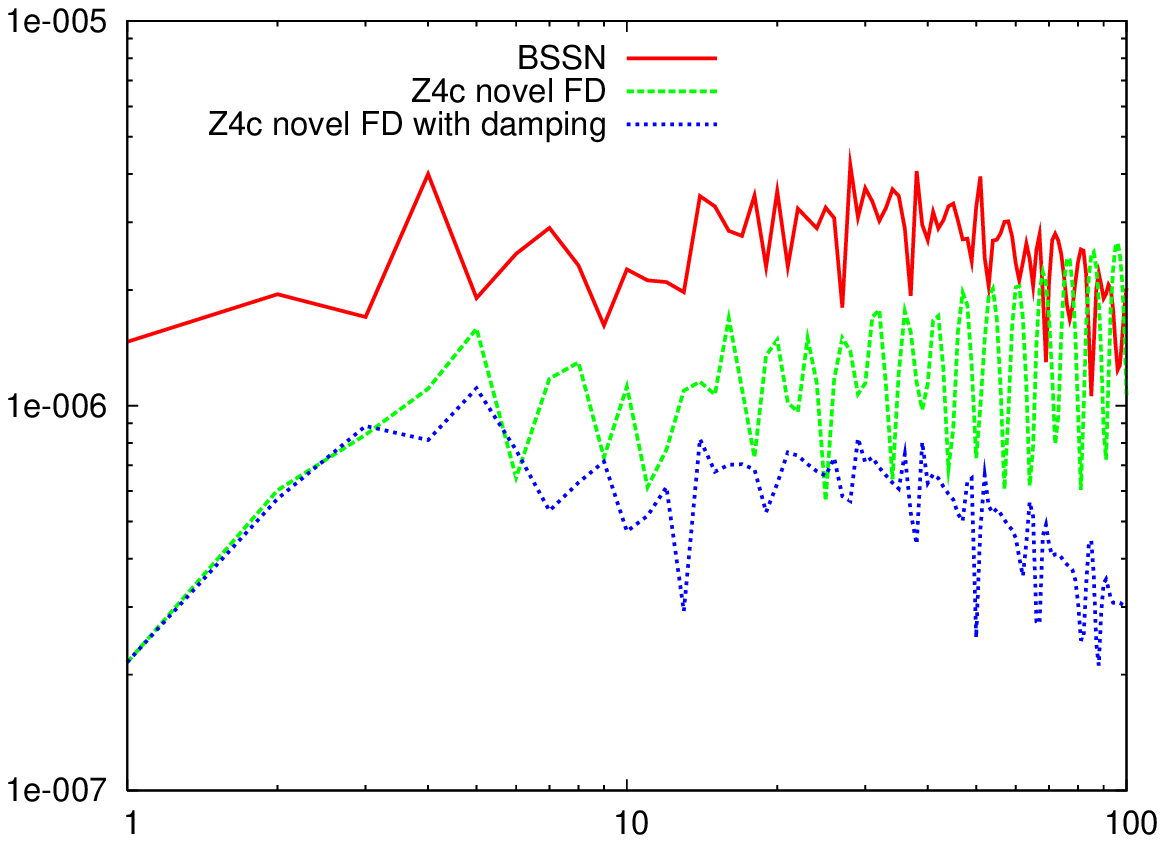}
\end{tabular}
\caption{ \label{fig:Gowdy_monitor} Expanding Gowdy wave test with
our puncture gauge condition. Courant factor $C=0.05$ is used. Left
subplot shows the $D_+$ norm of the solution difference between
resolution $\rho=4$ and $\rho=2$. Right subplot compares the
constraint monitor respect to time for BSSNOK formalism and Z4c
formalism. For Z4c formalism we only show novel finite difference
scheme here because standard finite difference scheme gives almost
the same result. Here for Z4c formalism, with ($\kappa_1=0.02$,
$\kappa_2=0$) and without damping term are also compared.}
\end{figure*}

The final tests are the expanding and collapsing Gowdy waves.
Following the suggestion of~\cite{AlcAllBon03,BabHusAli08}, we
choose for initial data
\begin{align}
&\gamma_{xx}=t^{-1/2}e^{\Lambda/2},\quad
\gamma_{yy}=te^{P},\quad
\gamma_{zz}=te^{-P},\nonumber\\
&K_{xx}=\pm\frac{1}{4}t^{-1/4}e^{\Lambda/4}(t^{-1}-\Lambda_{,t}),\quad
\alpha=t^{-1/4}e^{\Lambda/4},\nonumber\\
&K_{yy}=\pm\frac{1}{2}t^{1/4}e^{-\Lambda/4}e^P(-1-tP_{,t}),\nonumber\\
&K_{zz}=\pm\frac{1}{2}t^{1/4}e^{-\Lambda/4}e^{-P}(-1+tP_{,t}),\label{gowdyini}
\end{align}
where `+' sign is for the expanding Gowdy wave test and `-' sign is for
the collapsing Gowdy wave test, and other variables zero. Here
$P=J_0(2\pi t)\cos(2\pi x)$ and
\begin{align}
\Lambda&=-2\pi t J_0(2\pi t)J_1(2\pi t)\cos^2(2\pi x)\nonumber\\
&+2\pi^2t^2[J^2_0(2\pi t)+J^2_1(2\pi t)]\nonumber\\
&-\frac{1}{2}\big[(2\pi)^2[J^2_0(2\pi)+J^2_1(2\pi)]
-2\pi J_0(2\pi)J_1(2\pi)\big]
\end{align}
where $J_n$ are Bessel functions. Initially we set
$t=9.8753205829098$ following~\cite{AlcAllBon03,BabHusAli08}
exactly. The Courant factor is set as $C=0.05$. Here the use of  the
puncture gauge is troublesome because it does not reproduce the
preferred analytical gauge in the evolution. Unfortunately simply
choosing the lapse and shift apriori results in an ill-posed initial
value problem, highlighting the fact that in numerical relativity
there are still open questions related to the choice of gauge
conditions. Obviously conditions must be chosen that render the
initial value problem well-posed, but on the other hand such
conditions may not correspond to those best-suited to study a
problem analytically.

\begin{figure}[ht]
\begin{tabular}{c}
\includegraphics[width=0.5\textwidth]{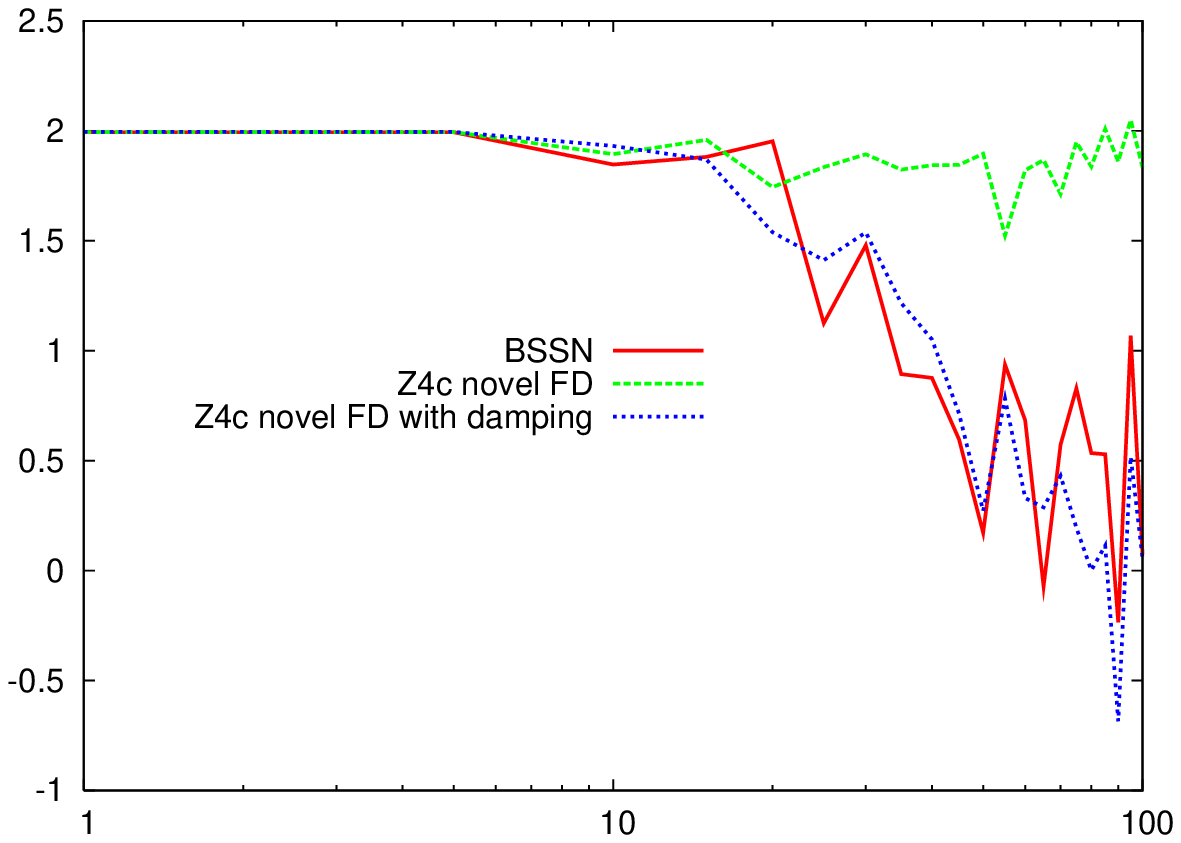}
\end{tabular}
\caption{ \label{fig:Gowdy_converge} Expanding Gowdy wave test with
our puncture gauge condition. Courant factor $C=0.05$ is used.
Convergence factor log${}_2(\frac{H(\rho=4)}{H(\rho=2)})$. Here
$H(\rho)$ means the Hamiltonian constraint violation for resolution
$\rho$.}
\end{figure}

The test results of expanding Gowdy wave are plotted in
Fig.~\ref{fig:Gowdy_monitor}. Although this test is very
tough~\cite{BabHusAli08}, upto~$1000$, neither the BSSNOK formalism
nor the Z4c formalism suffer from large growth of the errors. Our 
results demonstrate the robustness of puncture gauge condition. In 
the left subplot, we show the behavior of relative error for two 
different resolutions where~$\rho=2$ and~$\rho=4$ are used. From 
this result we can see the increasing behavior of the error roughly 
follows power function of~$t$ instead of the exponentially increasing 
result obtained by the spectral code in~\cite{BoyLinPfe06}. In the 
right subplot we show the constraint monitor with respect to time. In 
Fig.~\ref{fig:Gowdy_converge} we compare the convergence behavior 
with respect to Hamiltonian constraint violation. We find that Z4c 
formalism achieves better convergence results than BSSNOK formalism 
at this resolution. Once again the novel finite difference scheme 
gives almost exactly the same result as the standard discretization. 
If the damping term with~$\kappa_1=0.02$ and~$\kappa_2=0$ is used, 
the constraint violation is significantly reduced,
but, as shown in Fig.~\ref{fig:Gowdy_monitor}, does not reduce 
the~$D_+$ norm of the difference between two resolutions. 
Unfortunately in~Fig.\ref{fig:Gowdy_converge} we see that the damping 
term does reduce the experimental convergence factor.

\begin{figure*}[ht]
\begin{tabular}{cc}
\includegraphics[width=0.5\textwidth]{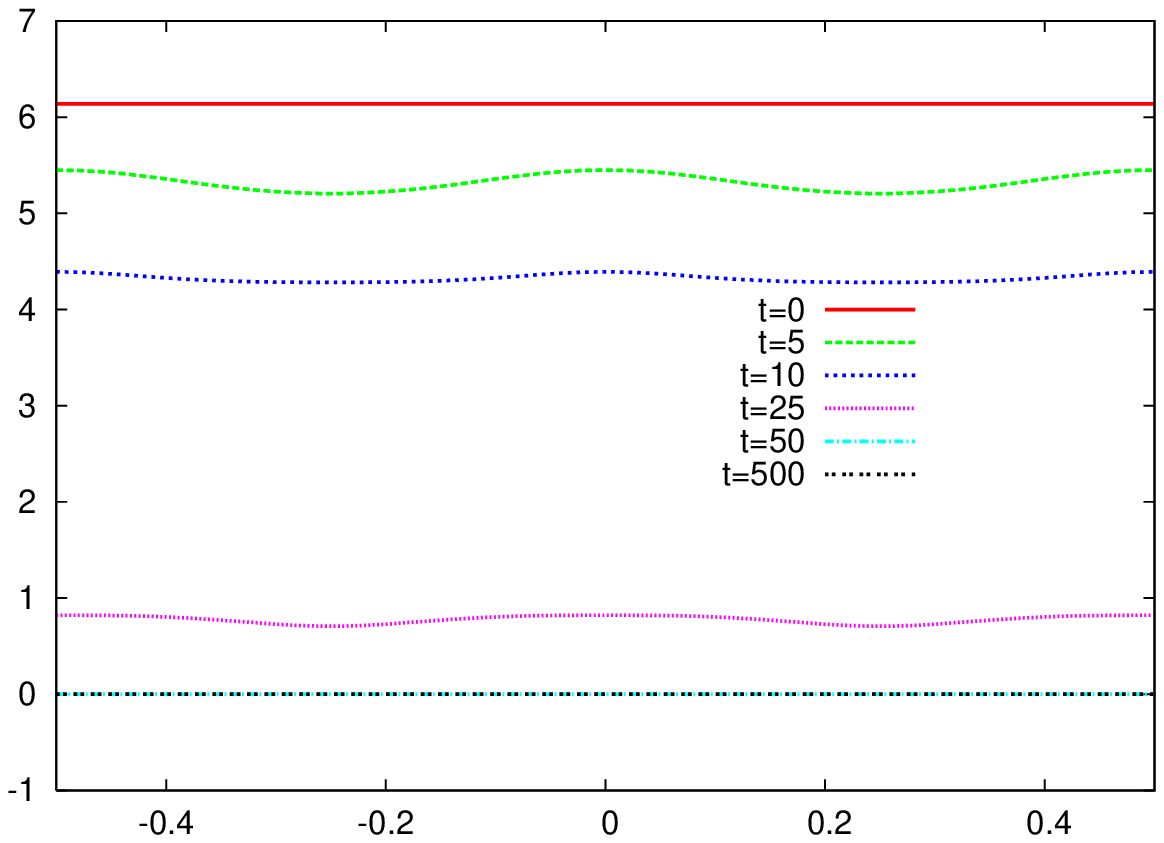}&
\includegraphics[width=0.5\textwidth]{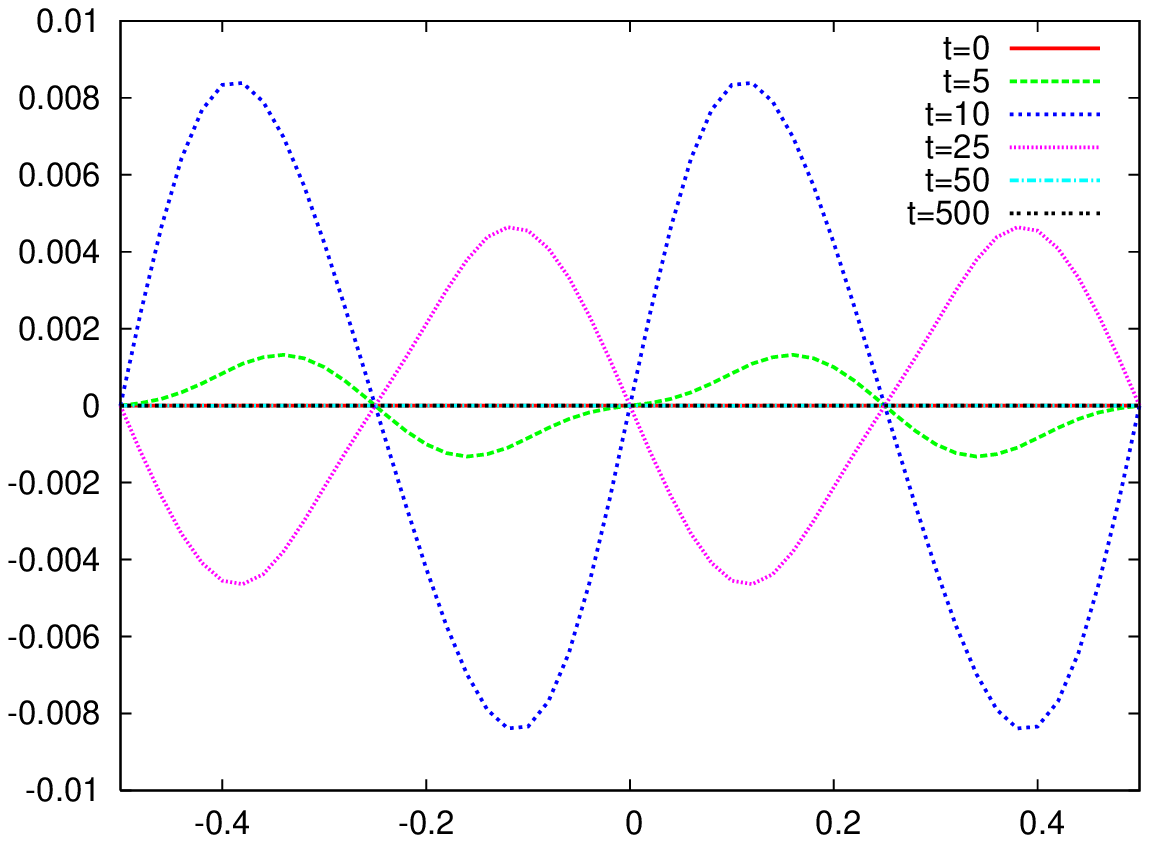}
\end{tabular}
\caption{ \label{fig:Gowdy_collapsing} Collapsing Gowdy wave test.
Courant factor is 0.05. BSSNOK formalism is used with resolution
$\rho=1$. Left subplot is the snapshot of $\alpha$ at different
time; right subplot is for $\beta^x$. $\alpha$ approaches 0 to
freeze the evolution. $\beta^x$ increases at very early time to
adapt the evolution and decrease to 0 also soon later. They together
make the evolution freeze near the universe singularity.}
\end{figure*}

For the collapsing Gowdy wave, in contrast to the proposal of
original ``{\it apples with apples}'' tests, we do not transform the
time coordinate to avoid singularity by hand. Instead we simply
change the sign of initial extrinsic curvature $K_{ij}$ in
Eq.(\ref{gowdyini}) and evolve with the puncture gauge. Until $1000$
light crossing times, when we stop the test, the code does not
crash. As the solution approaches the singularity, the puncture
gauge makes the evolution effectively stop by the collapse of the
lapse. Both BSSNOK and Z4c formalism gives almost identical results.
In Fig.~\ref{fig:Gowdy_collapsing} we show several snapshots for the
lapse and shift vector, and the singularity avoiding property of
puncture gauge condition can be clearly seen. Since BSSNOK and Z4c
formalisms give the similar result, only BSSNOK results are shown.
For the collapsing Gowdy test we find that the effect of the 
constraint damping scheme, with~$\kappa_1=0.02,\kappa_2=0$, is 
negligible.


\section{Conclusions}
\label{Section:Conclusions}


In order to demonstrate that a conformal decomposition of the Z4
formulation of general relativity may be a useful tool for numerical
relativity simulations in three spatial dimensions, we have studied
numerical stability of the system coupled to the moving puncture
gauge condition. First, we proved that in combination with a novel
discretization scheme, and periodic boundary conditions in space,
the Z4c system is numerically stable when linearized around a
constant background metric. We would like to extend our results to
the variable coefficient case, but since Z4 coupled to the puncture
gauge appears to be only strongly, but not symmetric hyperbolic, see
appendix~\ref{App:Not_Sym}, there are few methods to hand. We then
performed a complete set of the ``{\it apples with apples}'' tests,
which we modified slightly to take into account the status of the
initial value problem from the PDEs point of view, and discussed for
each the relative behavior of Z4c in comparison with BSSNOK. We
summarize the results of the tests as follows:

\paragraph*{Robust stability:} In the robust stability tests we
found that the Z4c evolutions result in a smaller constraint
violation, a feature that was insensitive to the choice of
gauge.

\paragraph*{Linear waves:} In the linear wave test we found only
marginal differences between the two formulations, regardless
of gauge conditions. It is perhaps expected that in this test
the differences will be small, because the propagation of
gravitational waves is of course a feature shared by both
systems.

\paragraph*{Gauge waves:} For the gauge waves test we found,
using the puncture gauge, that the Z4c evolutions exhibit an
experimentally computed self-convergence factor closer to the
expected second order than the BSSNOK data over many light
crossing times. In this context we found that the puncture
gauge has a strong symmetry seeking effect and rapidly pushes
the lapse to a constant, and the shift to zero.

\paragraph*{Gowdy waves:} To evolve expanding and collapsing
Gowdy spacetimes we again employed the puncture gauge. Here
we find that although neither Z4c or BSSNOK results in code
crashes, long-term convergence is very difficult to achieve.
Another problematic feature here is that although the
puncture gauge results in a well-posed initial value problems
for each of the formulations, it does not track the preferred
coordinates on the spacetime under consideration.

In appendix~\ref{App:weak} we also studied the effect of the
algebraic constraint projection on numerical convergence. When
the algebraic constraints are violated, the Z4c formulation is
only weakly hyperbolic, which is reflected in numerical
approximation as a failure to converge to the continuum
solution, even for trivial initial data such as noise on top
of the Minkowski spacetime.

The evidence for the usefulness of a conformal decomposition
of the Z4 formulation in astrophysical applications is mounting.
In spherical symmetry, the system was shown to have favourable
properties over BSSNOK~\cite{BerHil09}, especially in the evolution
of spacetimes containing matter. The trivial wave-like nature of
the constraint subsystem was then used to construct high-order
constraint preserving boundary conditions~\cite{RuiHilBer10},
which were studied in numerical applications in spherical
symmetry. The constraint damping scheme for the formulation was
studied in detail in~\cite{WeyBerHil11}. There, once again in
spherical symmetry, the efficacy of the constraint damping
scheme was studied in numerical applications and it was found
that the constraint damping scheme is a useful tool for the
suppression of violations in vacuum spacetimes, but that when
matter is present more work is required to paint a clear picture.
For the first time, numerical evolutions of binary black-hole
spacetimes with a conformal decomposition of Z4, CCZ4, were 
presented in~\cite{AliBonBon11}. Here we have focused on formal 
numerical stability, and compared numerical results obtained with 
BSSNOK and Z4c in a simple context. Overall we find some benefit 
to the use of Z4c, although in some sense the tests we have performed
here, namely the evolution of vacuum spacetimes with periodic
boundary conditions, are not optimally suited to the advantages
found in spherical symmetry. In our view two points remain to
be addressed before one can consider a whole-sale replacement
of BSSNOK in applications. First, it is desirable to obtain
boundary conditions that are constraint preserving, minimise
the incoming gravitational wave content, lead to a well-posed 
initial boundary value problem and can be implemented in a 3D 
production code. Second, conclusive evidence of the benefits of 
Z4c over BSSNOK is needed in applications. We hope to address 
both of these issues shortly.


\acknowledgements


The authors would like to thank Dana Alic, Sebastiano Bernuzzi, 
Carles Bona, Bernd Br\"ugmann, Carsten Gundlach, Ian Hinder,
Sascha Husa, Carlos Palenzuela, Luciano Rezzolla, Milton Ruiz, 
Yuichiro Sekiguchi and Andreas Weyhausen for helpful discussions, 
and, or, comments on the manuscript. This work was supported in 
part by DFG grant SFB/Transregio~7 ``Gravitational Wave Astronomy'' 
and by the NSFC (No.~10731080 and No.~11005149).


\appendix



\section{Z4c without enforcement of the algebraic constraints}
\label{App:weak}



\subsection{Weak hyperbolicity with the puncture gauge}


Without assuming that the algebraic constraints are enforced,
the equations of motion for the Z4c formulation can not
naturally be written in terms of the ADM variables. Under the
standard $2+1$ decomposition~\cite{GunGar05} and upto transverse and
non-principal derivatives, they are given by
{\allowdisplaybreaks
\begin{align}
\p_t\chi&=\frac{2}{3}\chi[\alpha(\hat{K}+2\Theta)-\p_s\beta^s]
+\beta^s\p_s\chi,\\
\p_t\tilde{\gamma}_{ss}&=-2\alpha\tilde{A}_{ss}
+\frac{4}{3}\chi\p_s\beta^s
+\beta^s\p_s\tilde{\gamma}_{ss},\\
\p_t\tilde{\gamma}_{qq}&=-2\alpha\tilde{A}_{qq}
-\frac{4}{3}\chi\p_s\beta^s
+\beta^s\p_s\tilde{\gamma}_{qq},\\
\p_t\alpha&= -\mu_L\alpha^2 \hat{K}+\beta^s\p_s\alpha,\\
\p_t\beta^s&= \alpha^2\mu_S\tilde{\Gamma}^s+\beta^s\p_s\beta^s,\\
\p_t\hat{K}&=-\p_s\p_s\alpha+\beta^s\p_s\hat{K},\\
\p_t\tilde{A}_{ss}&=\frac{\alpha}{3}\p_s\p_s\chi
-\frac{\alpha}{3}\p_s\p_s\tilde{\gamma}_{ss}
+\frac{\alpha}{6}\p_s\p_s\tilde{\gamma}_{qq}
-\frac{2}{3}\chi\p_s\p_s\alpha\nonumber\\
&+\frac{2}{3}\alpha\chi^2\p_s\tilde{\Gamma}^s
+\beta^s\p_s\tilde{A}_{ss},\\
\p_t\tilde{A}_{qq}&= -\frac{\alpha}{3}\p_s\p_s\chi
+\frac{\alpha}{3}\p_s\p_s\tilde{\gamma}_{ss}
-\frac{\alpha}{6}\p_s\p_s\tilde{\gamma}_{qq}
+\frac{2}{3}\chi\p_s\p_s\alpha\nonumber\\
&-\frac{2}{3}\alpha\chi^2\p_s\tilde{\Gamma}^s
+\beta^s\p_s\tilde{A}_{qq},\\
\p_t\Theta&=
\frac{\alpha}{\chi}\p_s\p_s\chi
-\frac{\alpha}{4\chi}\p_s\p_s\tilde{\gamma}_{ss}
-\frac{\alpha}{4\chi}\p_s\p_s\tilde{\gamma}_{qq}
+\frac{\alpha}{2}\chi\p_s\tilde{\Gamma}^s\nonumber\\
&+\beta^s\p_s\Theta,\\
\p_t\tilde{\Gamma}^s&=
\frac{4}{3\chi}\p_s\p_s\beta^s
-\frac{4\alpha}{3\chi}\p_s\hat{K}
-\frac{2\alpha}{3\chi}\p_s\Theta
+\beta^s\p_s\tilde{\Gamma}^s,
\end{align}}
in the scalar sector,
{\allowdisplaybreaks
\begin{align}
\p_t\tilde{\gamma}_{sA}&=-2\alpha\tilde{A}_{sA}
+\beta^s\p_s\tilde{\gamma}_{sA},\\
\p_t\beta^{A}&=\alpha^2\mu_S\tilde{\Gamma}^A+\beta^s\p_s\beta^A,\\
\p_t\tilde{A}_{sA}&=
-\frac{\alpha}{2\chi}\p_s\p_s\tilde{\gamma}_{sA}
+\frac{\alpha}{2}\chi\p_s\tilde{\Gamma}_A
+\beta^s\p_s\tilde{A}_{sA},\\
\p_t\tilde{\Gamma}^{A}&=
\frac{1}{\chi}\p_s\p_s\beta^A
+\beta^s\p_s\tilde{\Gamma}^A,
\end{align}
}
in the vector sector and
\begin{align}
\p_t\tilde{\gamma}_{AB}^{\textrm{TF}}&=-2\alpha\tilde{A}_{AB}^{\textrm{TF}}
+\beta^s\p_s\tilde{\gamma}_{AB}^{\textrm{TF}},\\
\p_t\tilde{A}_{AB}^{\textrm{TF}}&=
-\frac{\alpha}{2}\p_s\p_s\tilde{\gamma}_{AB}^{\textrm{TF}}
+\beta^s\p_s\tilde{A}_{AB}^{\textrm{TF}},
\end{align}
in the tensor sector. Here the analysis is essentially for the
linearized system, and we use the background metric~$\gamma_{ij}$
to raise and lower indices. We also use the convention that the
spatial vector~$s^i$ is normalized to one against the background
physical metric rather than using the conformal metric. The
principal symbol of each sector can be trivially read off from
these equations. In geometric units the tensor sector has
speeds~$\pm 1$, and a complete set of characteristic variables.
Likewise the vector sector has speeds~$\pm (1,\sqrt{\bar{\mu}_S})$,
with~$\bar{\mu}_S=\mu_S/\chi$ and a complete set of characteristic
variables. The scalar sector has speeds~$0,\pm(1,\sqrt{\mu_L},
\sqrt{4\bar{\mu}_S/3})$, but does not have a full set of
characteristic variables; the formulation is thus only weakly
hyperbolic, and so does not have a well-posed initial value
problem. It may be possible to modify the equations of motion
by adding (possibly derivatives) of the algebraic
constraints~$D$ and $T$ to the equations of motion to achieve
strong, or even symmetric hyperbolicity. But in any case it
is not true that when the algebraic constraints are violated
the PDE properties of the system are unaltered.


\subsection{Numerical convergence without constraint
projection}


\begin{figure*}[t]
\begin{tabular}{cc}
\includegraphics[width=0.5\textwidth]{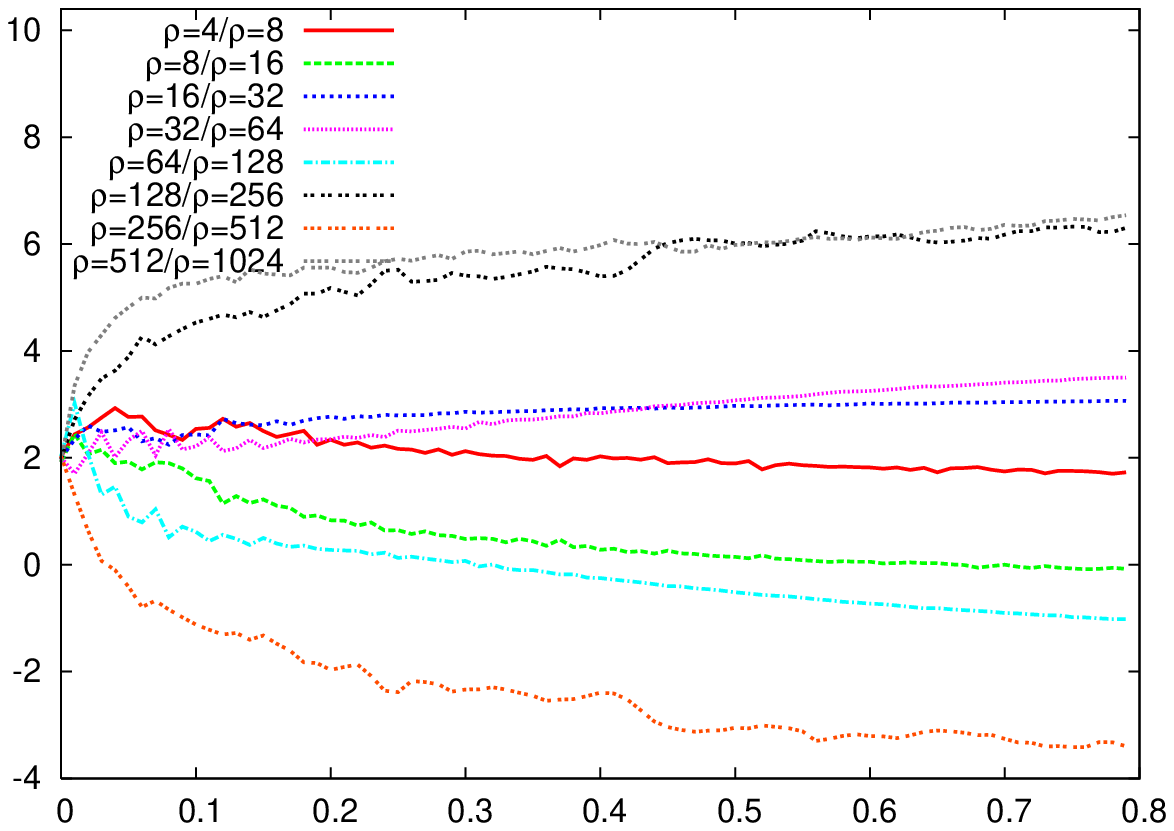}&
\includegraphics[width=0.5\textwidth]{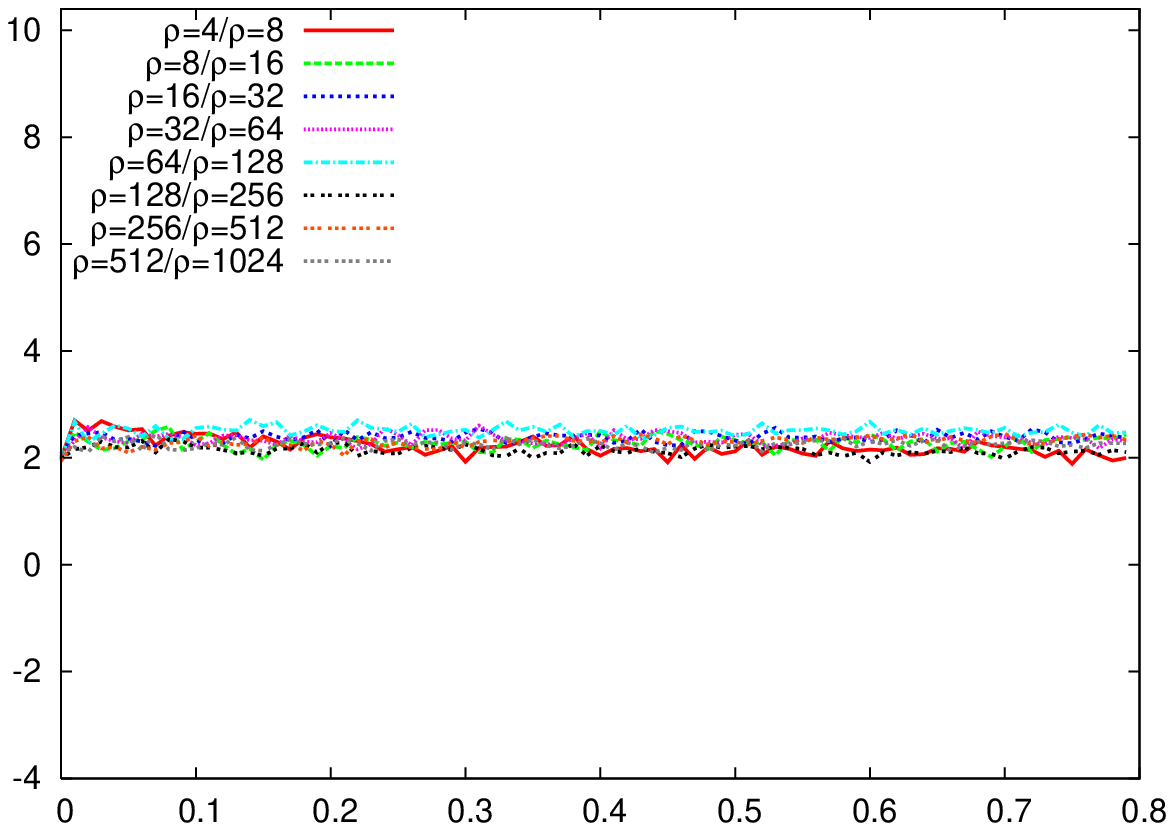}
\end{tabular}
\caption{
\label{fig:Projection}
Convergence test with the puncture gauge
condition~(\ref{eqn:BM_lapse}-\ref{eqn:Gamma_driver}). Convergence
factors~\eqref{eqn:confactor} in time for different
resolution are shown. The left subplots are without algebra
constraint enforcement while the right ones with. Here the
adjusted finite difference scheme is used.}
\end{figure*}

In the previous subsection we saw that when the algebraic
constraints are not enforced, the Z4c formulation is only
weakly hyperbolic. The open question is how ill-posedness
of the initial value problem will be reflected in numerical
approximation. To answer the question we follow~\cite{Hin05}
and perform convergence tests, using the~$D_+$ norm, as
defined by~\eqref{eqn:D_+norm}, using initial data inspired
by the robust stability test. Specifically we set the initial
perturbation
to~$\epsilon\in(-10^{-3}/\rho^p,10^{-3}/\rho^p)$. For the metric
components we set~$p=3$, and use $p=2$ for the remaining
variables. This choice is needed to guarantee second order
consistency of the initial data in the~$D_+$ norm, which is of
course necessary for convergence in that norm. We employ the
same numerical grid used in the robust stability tests, but
instead of evolving for of~$1000$ light crossing times, we
consider the interval~$t\in[0,0.8]$. We set the Courant
factor to~$\lambda=0.5$, and use artificial
dissipation~\eqref{eqn:KO_diss} as usual with~$\sigma=0.02$.
We consider higher resolutions
than those used in the robust stabiliy test, and compute
additionally data with~$\rho=16,\dots, 1024,$ in factors of
two. Before calculating the the $D_+$ norm, we restrict the
solution to the~$\rho=1$ grid. The convergence factor with
respect to resolutions~$\rho$ and~$2\rho$ is
\begin{eqnarray}
\label{eqn:confactor}
{\cal C}\equiv\log_2
\left(\frac{||u_{\rho}-u_{\textrm{exact}}||_{D_+}}
{||u_{2\rho}-u_{\textrm{exact}}||_{D_+}}\right).
\end{eqnarray}
The convergence factors obtained with the puncture
gauge condition~(\ref{eqn:BM_lapse}-\ref{eqn:Gamma_driver})
and the novel discretization are shown in
Fig.~\ref{fig:Projection}. Very similar results are obtained
with the standard discretiazation and we interpret the close
agreement of the two tests as numerical evidence that the
standard discretization is formally numerically stable, with
constraint projection, despite our inablility to prove so.
At the lowest three resolutions the constraint projection does
not seem to have any effect on the result, but at higher
resolutions the conclusion is clear: without algebraic
constraint projection the numerical solution
does not converge to the continuum solution. Even with the
constraint projection we do not seem to obtain perfect
second order convergence, but since we are evolving random
noise and the trend is clear, we conclude that with constraint
projection the scheme is converging. Furthermore in our tests
with smooth initial data we do obtain perfect second order
convergence with the constraint projection. We obtain similar
results with a Courant factor $\lambda=0.1$, and with the harmonic
slicing and zero shift. We also find similar results when
discretizing at fourth order.


\section{Symmetric Hyperbolicity of Z4 with the puncture
gauge}\label{App:Not_Sym}


\paragraph*{Definitions:} A quantity~$E$ conserved by the
principal part of the second order system with state
vector~$u$,
\begin{align}
E&=\int \epsilon \textrm{d}x,\nonumber\\
\epsilon&\equiv (\p_ju,\p_tu)H^{ij}(\p_iu,\p_tu)^\dagger,
\end{align}
is called a candidate energy. The Hermitian matrix~$H$ is
called a candidate symmetrizer. A system that admits a
positive definite candidate symmetrizer is called symmetric
hyperbolic. This definition is equivalent to the existence of
a first order reduction of the system that is symmetric
hyperbolic according to the standard definition for first order
systems.

\paragraph*{Z4c second order form:} The principal part of the
equations of motion of Z4c coupled to the puncture gauge can
be written
\begin{align}
\p^2_0\alpha&=\mu_L\gamma^{ij}\p_i\p_j\alpha,\\
\p^2_0\beta^i&=\bar{\mu}_S\gamma^{jk}\p_j\p_k\beta^i
+\frac{\bar{\mu}_S}{\mu_L}\p^i\p_0\alpha
+\frac{1}{6}\alpha\bar{\mu}_S\gamma^{jk}\p^i\p_0\gamma_{jk},\\
\p^2_0\gamma_{ij}&=\gamma^{kl}\p_k\p_l\gamma_{ij}
+\frac{1}{3}\gamma^{kl}\p_i\p_j\gamma_{kl}
+\frac{2}{\alpha}\p_i\p_j\alpha\nonumber\\
&+\frac{2}{\alpha}\Big(1-\frac{1}{\bar{\mu}_S}\Big)
\gamma_{k(i}\p_{j)}\p_0\beta^k,
\end{align}
where we write~$\p_0=(\p_t-\beta^i\p_i)/\alpha$. We thus express
the principal part of the system in the form
\begin{align}
\p_0u_{i\,kl}&=\mathfrak{A}^p{}_i{}^j{}_{kl}{}^{mn}\p_pu_{j\,mn},
\end{align}
with
\begin{align}
\p_0u_{i\,kl}&=(\p_i\gamma_{kl},\p_i\alpha,\p_i\beta_k,
\p_0\gamma_{kl},\p_0\alpha,\p_0\beta_k)^\dagger,
\end{align}
and the principal part matrix given by
\begin{align}
\mathfrak{A}^p{}_i{}^j{}_{kl}{}^{mn}\p_pu_{j\,mn}&=
\left(\begin{array}{cc}
 0 & \delta^p{}_iI_{kl}{}^{mn}\\
 \mathcal{A}^{pj}{}_{kl}{}^{mn}  & \mathcal{B}^{p}{}_{kl}{}^{mn}
\end{array}\right),
\end{align}
with~$I_{kl}{}^{mn}$ the appropriate identity and
\begin{align}
\mathcal{A}^{pj}{}_{kl}{}^{mn}&=
\left(\begin{array}{ccc}
\mathcal{A}_{11}^{pj}{}_{kl}{}^{mn} &
\frac{2}{\alpha}\delta^p{}_{(k}\delta^j{}_{l)} & 0\\
0 & \mu_L\gamma^{pj} & 0\\
0 & 0 & \bar{\mu}_S\gamma^{pj}\delta^k{}_m
\end{array}\right),\\
\mathcal{B}^{p}{}_{kl}{}^{mn}&=
\left(\begin{array}{ccc}
0 & 0 &
\frac{2}{\alpha}\big(1-\frac{1}{\bar{\mu}_S}\big)
\delta^p{}_{(k}\gamma{}_{l)m}\\
0 & 0 & 0\\
\frac{\alpha}{6}\bar{\mu}_S\gamma^{mn}\gamma^{pk} &
\frac{\bar{\mu}_S}{\mu_L}\gamma^{pk} & 0
\end{array}\right),
\end{align}
where
\begin{align}
\mathcal{A}_{11}^{pj}{}_{kl}{}^{mn}&=
\gamma^{pj}\delta^m{}_{(k}\delta^n{}_{l)}
+\frac{1}{3}\gamma^{mn}\delta^p{}_{(k}\delta^j{}_{l)}.
\end{align}

\paragraph*{Ansatz candidate and conservation:} Starting from
the ansatz candidate~\cite{HilRic10},
\begin{align}
\label{eqn:candidate}
&H^{ij\,kl\,mn}=\nonumber\\
&\left(
\begin{array}{cccccc}
H_{11}^{ij\,kl\,mn} & H_{12}^{ij\,kl} & 0 & 0 & 0 & H_{16}^{i\,kl\,m}\\
H_{12}^{ji\,mn} & H_{22}^{ij} & 0 & 0 & 0 & H_{26}^{i\,m}\\
0 & 0 & H_{33}^{ij\,k\,m} & H_{34}^{i\,k\,mn} & H_{35}^{i\,k} & 0\\
0 & 0 & H_{34}^{j\,m\,kl} & H_{44}^{kl\,mn} & H_{45}^{kl} & 0\\
0 & 0 & H_{35}^{j\,m} & H_{45}^{mn} & H_{55} & 0\\
H_{16}^{j\,mn\,k} & H_{26}^{j\,k} & 0 & 0 & 0 & H_{66}^{k\,m}
\end{array}\right),
\end{align}
where we define
{\allowdisplaybreaks
\begin{align*}
H_{11}^{ij\,kl\,mn} &=
H_{11}^1 \gamma^{ij} \gamma^{kl} \gamma^{mn}+
2 H_{11}^2 \gamma^{ij} \gamma^{k(m} \gamma^{n)l}\\
&\qquad +
2 H_{11}^3 (\gamma^{i(k} \gamma^{l)j} \gamma^{mn} +
            \gamma^{i(m} \gamma^{n)j} \gamma^{kl})\\
&\qquad +
2 H_{11}^4 (\gamma^{ik} \gamma^{j(m} \gamma^{n)l} +
            \gamma^{il} \gamma^{j(m} \gamma^{n)k}\\
&\qquad\qquad + \gamma^{im} \gamma^{j(k} \gamma^{l)n} +
            \gamma^{in} \gamma^{j(k} \gamma^{l)m})\\
&\qquad +
A_{11}^1 (\gamma^{k[i}\gamma^{j](m}\gamma^{n)l} +
          \gamma^{l[i}\gamma^{j](m}\gamma^{n)m}),\\
H_{12}^{ij\,kl} &=
2 H_{12}^1 \gamma^{i(k} \gamma^{l)j} +
  H_{12}^2 \gamma^{ij} \gamma^{kl},\\
H_{22}^{ij} &= H_{22}^1 \gamma^{ij},\\
H_{33}^{ij\,k\,m} &=
2 H_{33}^1 \gamma^{i(k}\gamma^{m)j} +
H_{33}^2 \gamma^{ij} \gamma^{km}
+ A_{33}^1 \gamma^{k[i}\gamma^{j]m},\\
H_{16}^{i\,kl\,m} &=
2 H_{16}^1 \gamma^{i(k} \gamma^{l)m} +
H_{16}^2 \gamma^{im} \gamma^{kl},\\
H_{26}^{i\,m} &= H_{26}^1 \gamma^{im},\\
H_{34}^{i\,k\,mn} &=
H_{34}^1 \gamma^{i(m} \gamma^{n)k} +
H_{34}^2 \gamma^{ik} \gamma^{mn},\\
H_{35}^{i\,k} &= H_{35}^1 \gamma^{ik},\\
H_{44}^{kl\,mn} &=
2 H_{44}^1 \gamma^{k(m} \gamma^{n)l} +
H_{44}^2 \gamma^{kl} \gamma^{mn},\\
H_{45}^{kl} &= H_{45}^1 \gamma^{kl},\\
H_{66}^{k\,m} &= H_{66}^1 \gamma^{km}.
\end{align*}}
Because of the structure of the equations of motion and our
choice of gauge the block structure of the ansatz candidate is no
restriction. The restriction to energy densities constructed by
using the metric to constract indices does restrict the class
of symmetrizers, but is the largest natural choice. Conservation
of the energy is guaranteed by Hermicity of the matrix, suppressing
non-derivative indices,
\begin{align}
S_iH^{ij}\mathfrak{A}^p{}_j{}^ks_pS_k,
\end{align}
for every spatial vector $s^i$, where
\begin{align}
S_i&=
\left(\begin{array}{cc}
s_i & 0\\
0 & 1
\end{array}\right),
\end{align}
and the partition of this matrix is compatible with that
of~$u_{i\,kl}$ into spatial and time derivatives. Imposing
energy conservation for the puncture gauge implies that
\begin{align*}
H^1_{11}&=\frac{2}{9}(6 H^2_{44} - \alpha H^2_{34}),&\quad
H^2_{11}&=0,\\
H^3_{11}&=0, &\quad H^4_{11}&=0,\\
H_{12}^1&=0,  &\quad H_{16}^1&=0,\\
H_{16}^2&=\frac{4}{3\bar{\mu}_S}H_{34}^2, &\quad
H_{33}^1&=\frac{1-\bar{\mu}_S}{\bar{\mu}_S\alpha}H_{34}^2,\\
H_{34}^1&=0, &\quad H_{44}^1&=0,
\end{align*}
and the more complicated
{\allowdisplaybreaks
\begin{align*}
H^1_{22}&=\mu_LH_{55}
+\frac{8H_{34}^2(2\bar{\mu}_S-3)}
{\mu_L(3\mu_L-4\bar{\mu}_S)\alpha}
-\frac{12H_{44}^2(4+\mu_L-4\bar{\mu}_S}
{\mu_L(3\mu_L-4\bar{\mu}_S)\alpha^2},\\
H_{26}^1&=\frac{24(\bar{\mu}_S-1)}{\bar{\mu}_S(4\bar{\mu}_S-3\mu_L)}
(2H_{44}^2+\alpha H_{34}^2),\\
H^1_{35}&=\frac{24-6\mu_L-16\bar{\mu}_S}
{\mu_L(3\mu_L-4\bar{\mu}_S)\alpha}
H_{34}^2
+\frac{48(\bar{\mu}_S-1)}{\mu_L(4\bar{\mu}_S-3\mu_L)\alpha^2}
H_{44}^2,\\
H_{45}^1&=\frac{4\bar{\mu}_S}{\mu_L(4\bar{\mu}_S-3\mu_L)}H_{34}^2
+\frac{6}{4\bar{\mu}_S-3\mu_L}H_{44}^2,\\
H_{66}^1&=\frac{2(3\bar{\mu}_S-4)}{\bar{\mu}_S^2\alpha}H_{34}^2
+\frac{12(\bar{\mu}_S-1)}{\bar{\mu}_S^2\alpha^2}H_{44}^2,\\
H_{12}^2&=\frac{4}{4\bar{\mu}_S-3\mu_L}H_{34}^2+
\frac{8}{(4\bar{\mu}_S-3\mu_L)\alpha}H_{44}^2,\\
H_{33}^2&=\frac{2(3\bar{\mu}_S-4)}{\bar{\mu}_S\alpha} H_{34}^2
+\frac{12(\bar{\mu}_S-1)}{\bar{\mu}_S\alpha^2}H_{44}^2,
\end{align*}}
Note that the the terms antisymmetric in derivative indices
drop out of the conservation equations.

\paragraph*{Positivity:} Introducing and expanding with a tensor
basis reveals vanishing elements on the diagonal of the candidate
symmetrizer; there is no positive definite candidate symmetrizer
starting from the natural ansatz~\eqref{eqn:candidate}. We
thus conclude, unfortunately, that the use of summation by parts
finite differences for the Z4c formulation with the puncture
gauge will not guarantee numerical stability.


\bibliographystyle{apsrev}
\bibliography{refs}


\end{document}